\documentclass[journal]{IEEEtran}
\usepackage{cite}

\ifCLASSINFOpdf
\else
  \usepackage[dvips]{graphicx}
  \usepackage{subfigure}
\fi
\begin{document}
%
\title{High-Performance Complementary III-V Tunnel FETs with Strain Engineering}
%
%
%

\author{Jun~Z.~Huang,
        Yu Wang,
        Pengyu~Long,
        Yaohua~Tan,
        Michael~Povolotskyi,
        and~Gerhard~Klimeck
\thanks{J. Z. Huang, Y. Wang, P. Long, M. Povolotskyi, and G. Klimeck are with
the Network for Computational Nanotechnology and Birck Nanotechnology
Center, Purdue University, West Lafayette, IN 47907 USA (e-mail:
junhuang1021@gmail.com).}
\thanks{Y. Tan is with Department of Electrical and Computer Engineering,
University of Virginia, Charlottesville, VA 22904-4743 USA}
\thanks{Manuscript received xxx xx, 2016; revised xxxx xx, 2016. This work uses nanoHUB.org
computational resources operated by the Network for Computational
Nanotechnology funded by the U.S. National Science Foundation
under Grant EEC-0228390, Grant EEC-1227110, Grant EEC-0634750,
Grant OCI-0438246, Grant OCI-0832623, and Grant OCI-0721680. This
material is based upon work supported by the National Science Foundation
under Grant 1125017. NEMO5 developments were critically supported by an
NSF Peta-Apps award OCI-0749140 and by Intel Corp.}}

\maketitle

\begin{abstract}
Strain engineering has recently been explored to improve tunnel field-effect transistors (TFETs). Here, we report design and performance of strained ultra-thin-body (UTB) III-V TFETs by quantum transport simulations. It is found that for an InAs UTB confined in [001] orientation, uniaxial compressive strain in [100] or [110] orientation shrinks the band gap meanwhile reduces (increases) transport (transverse) effective masses. Thus it improves the ON state current of both n-type and p-type UTB InAs TFETs without lowering the source density of states. Applying the strain locally in the source region makes further improvements by suppressing the OFF state leakage. For p-type TFETs, the locally strained area can be extended into the channel to form a quantum well, giving rise to even larger ON state current that is comparable to the n-type ones. Therefore strain engineering is a promising option for improving complementary circuits based on UTB III-V TFETs.
\end{abstract}


%
\IEEEpeerreviewmaketitle

\section{Introduction}
%
%
%
%
\IEEEPARstart{A}{tunnel} field-effect transistor (TFET) is a steep subthreshold swing (SS) device that is promising in building future low-power integrated circuits. But its drive current ($I_{\rm{ON}}$) is usually limited by the small tunnel probability \cite{ionescu2011tunnel} leading to pronounced switching delay (CV/I). Various methods have been proposed to improve $I_{\rm{ON}}$, such as the doping engineering \cite{jhaveri2011effect,huang2016drc}, different channel materials (including low band gap III-V materials and two-dimensional materials) \cite{luisier2009performance}, broken/staggered gap heterojunction \cite{luisier2009performance,mohata2011demonstration}, grading of the source \cite{brocard2014large}, resonant enhancement \cite{avci2013heterojunction,pala2015exploiting,Long2016design}, and channel/source heterojunctions \cite{Long2016,Long2016drc}.

As a widely used technique to engineer electronic devices, strain has also been employed to improve the performance of silicon/germanium TFETs \cite{nayfeh2008design,Krishnamohan2008double,Boucart2009lateral} and III-V materials (InAs and InAs/GaSb) based nanowire TFETs \cite{conzatti2011simulation,brocard2013design,Huang2015,visciarelli2016impact}. These studies show that strain can reduce the band gap and/or effective masses leading to improved tunnel probability. At the same time, however, there are two side effects. One is that the direct source-to-drain tunneling and ambipolar leakage are also increased. The other is that the valence band density of states (DOS) is reduced, creating large source Fermi degeneracy (the distance between Fermi level and valence band edge) that degrades the SS toward thermal limit (60mV/dec). The first one can be mitigated by using local strain, i.e., by applying strain around the source only so that the band gap and effective masses in the channel remains unchanged \cite{Boucart2009lateral,Conzatti2013}. The second one has recently been addressed by lowering the doping concentration in the source but maintaining a high doping region next to the tunnel junction to shorten the tunnel distance \cite{Verreck2016}.

However, for experimentally more favorable ultra-thin-body (UTB) III-V TFETs, strain engineering has not been studied yet. UTB structures offer more design freedoms. For example, by selecting $(\bar{1}10)$/[110] instead of (001)/[100] as the confinement/transport orientation, the tunnel probability of GaSb/InAs heterojunction is increased due to the smaller band gap and transport effective masses \cite{Long2016}. Besides, the effect of strain is different if strain is applied in different directions \cite{Moussavou2015}.

Moreover, the efforts mentioned above mainly focus on strained n-type TFETs (nTFETs). Strain effects on p-type TFETs (pTFETs) still need to be explored as strain influences conduction and valence bands differently. In fact, due to the asymmetric conduction and valence band structures of III-V materials, pTFETs behave differently from nTFETs. For pTFETs, since the conduction band DOS is lower than valence band DOS, the allowed source doping density in pTFETs is lower than that of nTFETs. This leads to longer source depletion region and consequently smaller $I_{\rm{ON}}$ \cite{Avci2011,Avci2015}. The study in \cite{Verreck2014} shows that doping or heterojunction design in the source can help to solve this problem.

In this paper, we systematically study globally and locally strained InAs UTB nTFETs and pTFETs using accurate quantum transport simulations. Different types of strain and crystal orientations are investigated. It is found that, by selecting [001]/[100] orientation as the confinement/transport direction and applying uniaxial compressive strain along the transport direction, $I_{\rm{ON}}$ of both nTFETs and pTFETs can be significantly improved. Furthermore, the improvement is more pronounced for pTFETs, narrowing the performance gap between nTFETs and pTFETs. These make strain engineering a promising technique for enabling low-power high-performance complementary circuits based on III-V UTB TFETs.

\begin{table}
\caption{Material parameters for InAs at T=300K.}
\label{tab:parameters}       
\begin{tabular}{p{1cm}p{1cm}p{1cm}p{0.8cm}p{0.7cm}p{0.7cm}p{0.7cm}}
\hline\noalign{\smallskip}
$E_g$ (eV) & $\Delta$ (eV) & $E_p$ (eV) &$m_e^*/m_0$ &$\gamma_1$ & $\gamma_2$ & $\gamma_3$ \\
\noalign{\smallskip}
0.347  & 0.397    & 19.20  & 0.022    & 20.63     & 8.917   &  9.724   \\
\hline\noalign{\smallskip}
$a_v$ (eV) & $a_c$ (eV) & $b$ (eV) & $d$ (eV) & $C_{11}$ (GPa) & $C_{12}$ (GPa)  & $C_{44}$ (GPa) \\
\noalign{\smallskip}
-1.00  & -5.08    & -1.8    & -3.6     & 832.9   & 452.6   & 395.9 \\
\hline\noalign{\smallskip}
\end{tabular}
\end{table}

\section{Simulation Method}
The devices are simulated using NEMO5 tool \cite {Steiger2011} with Poisson equation and quantum ballistic transport equations (quantum transmitting boundary method \cite{Lent1990}) solved self-consistently. Phonon scattering is excluded in this work since it does not significantly impact the I-V characteristics of III-V homojunction TFETs \cite{luisier2010simulation,conzatti2011simulation}. As a widely used method to study strained III-V materials, the eight-band $\mathbf{k}\cdot\mathbf{p}$ method \cite{Bahder90} is employed to obtain the device Hamiltonian. It is also computationally more efficient than atomistic full-band tight binding (TB) method especially when strain is included.

\begin{figure}[htbp] \centering
{\includegraphics[width=8.7cm]{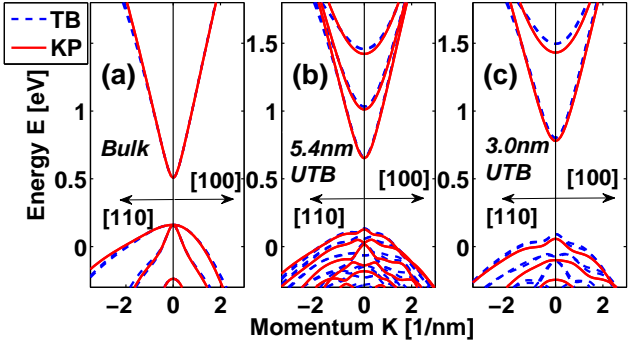}}
\caption{TB and KP calculations of $E$-$k$ diagram for (a) bulk InAs, (b) a 5.4nm thick UTB InAs, and (c) a 3.0nm thick UTB InAs. Both UTBs are confined in the [001] direction.}
\label{fig:ek}
\end{figure}

Since tunneling current is very sensitive to band structures, the accuracy of $\bf{k}\cdot\bf{p}$ method needs to be validated. We first extract the $\mathbf{k}\cdot\mathbf{p}$ band parameters from corresponding TB ($\rm{sp^3d^5s^*}$ basis with spin-orbit coupling) calculation of bulk material. The TB parameters taken from \cite{Tan2016} are fit to first-principles density functional theory (DFT) band structures as well as wave functions, and have been shown to be transferrable to UTB structures \cite{Tan2015}. The extraction procedure can be found in \cite{Huang2015}. The extracted $\mathbf{k}\cdot\mathbf{p}$ parameters for InAs used in this study are list in Table \ref{tab:parameters}. Note that the values of deformation potentials and elastic constants (for calculating Poisson ratio and converting stress to strain) are suggested in \cite{Vurgaftman01}. As shown in Fig. \ref{fig:ek} (a), for bulk InAs, the two calculations match well around $k=0$ as expected.

\begin{figure}[htbp] \centering
{\includegraphics[width=4.35cm]{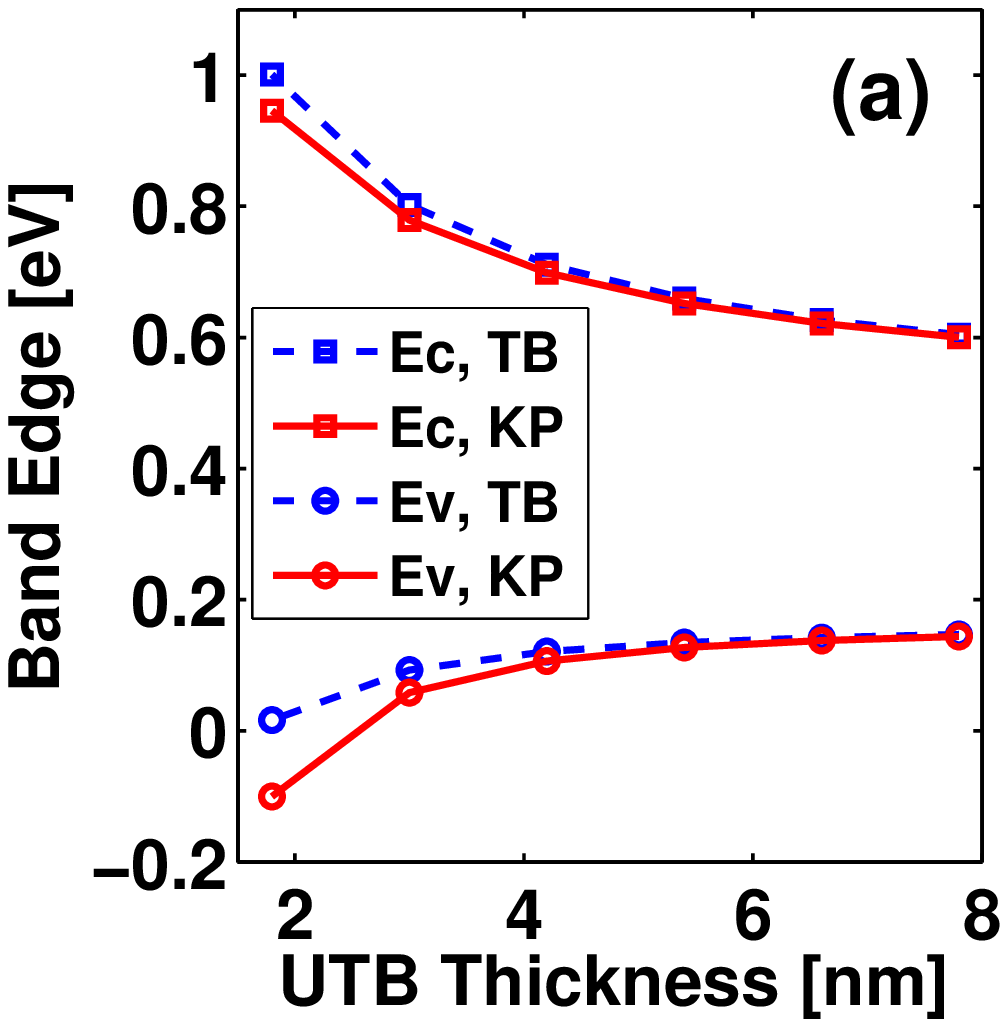}}
{\includegraphics[width=4.35cm]{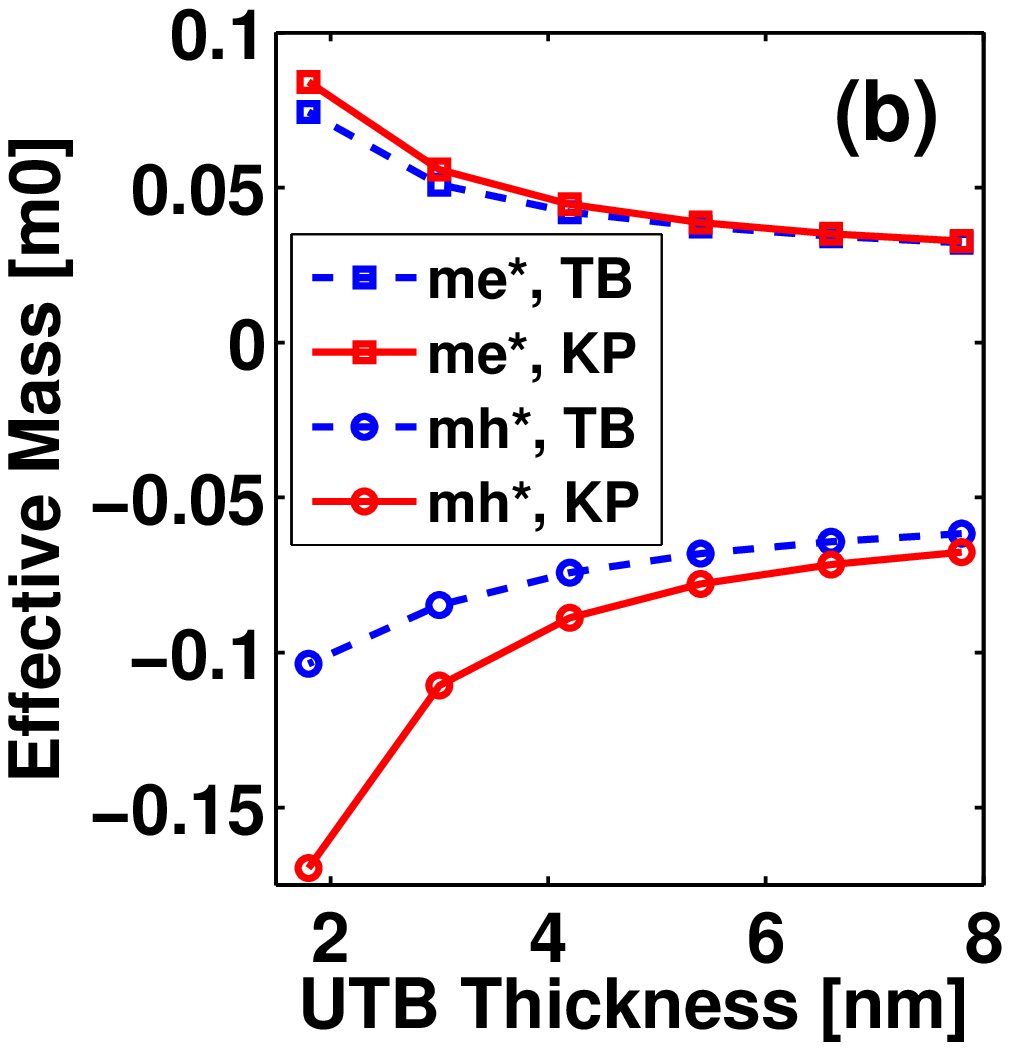}}
\caption{TB and KP calculations of (a) conduction band edge (Ec) and valence band edge (Ev), (b) electron effective mass (me*) and hole effective mass (mh*) in the [100] direction, for different InAs UTB thicknesses. The UTBs are confined in the [001] direction.}
\label{fig:band_edge_mass}
\end{figure}

We then compare the confined band structures of InAs UTBs calculated from both methods, as shown in Fig. \ref{fig:ek} (b) and (c). It is observed that the shapes of the band structures are quite similar to each other around the band gap, for both UTB thicknesses. The band edges and effective masses (which are of great interest for TFETs) for different UTB thicknesses are further calculated and plotted in Fig. \ref{fig:band_edge_mass}. It is seen that the $\bf{k}\cdot\bf{p}$ calculations match TB calculations quite well when the UTB is thick. When the UTB becomes thinner, the results start to diverge, especially for the hole effective mass. The general trend, i.e., the band gap and effective masses increase as UTB thickness decreases, is captured by both methods.

\section{Strained InAs UTBs}
\begin{figure*}[htbp] \centering
\subfigure[]{\includegraphics[width=5.8cm]{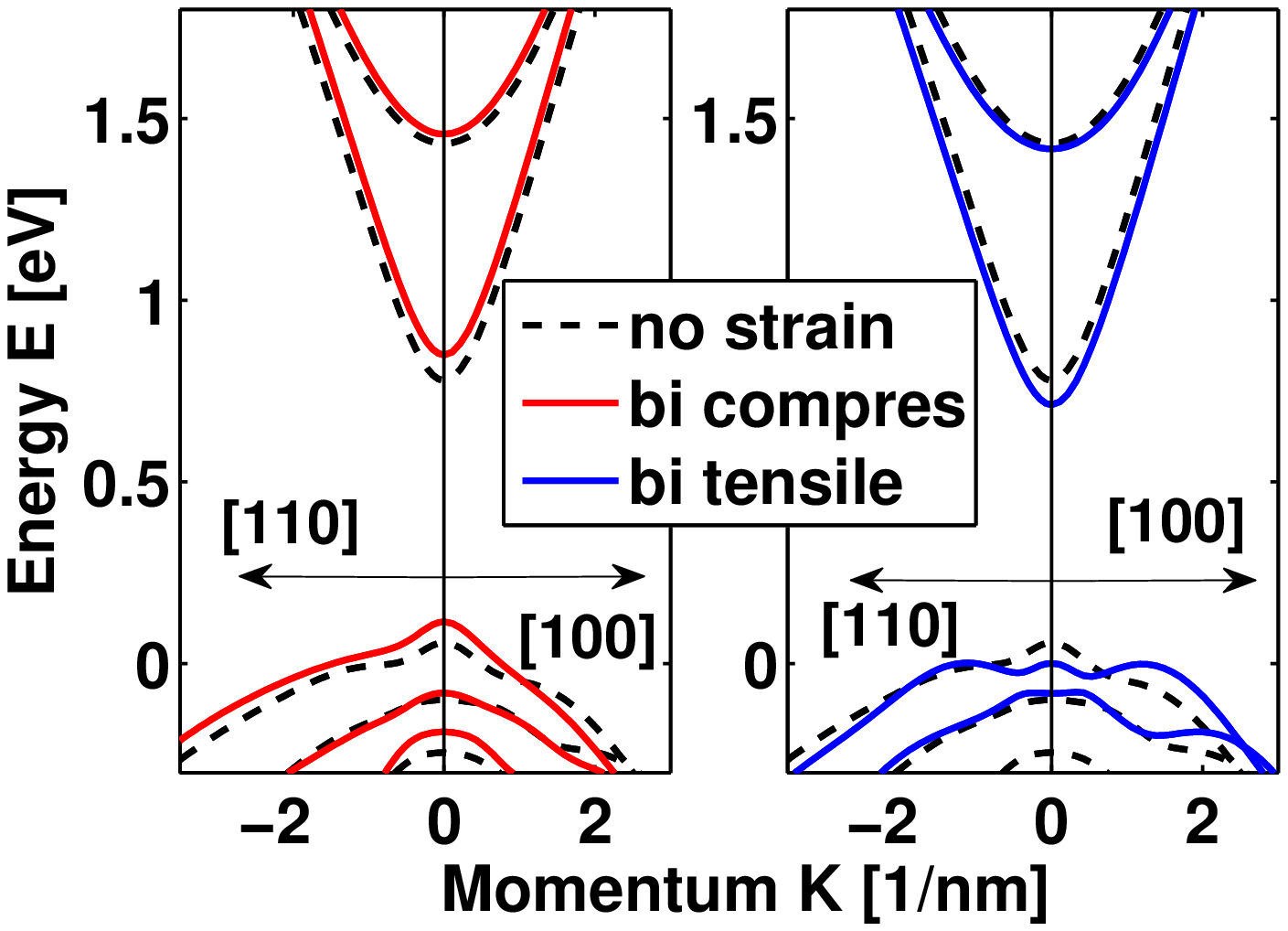}}
\subfigure[]{\includegraphics[width=5.8cm]{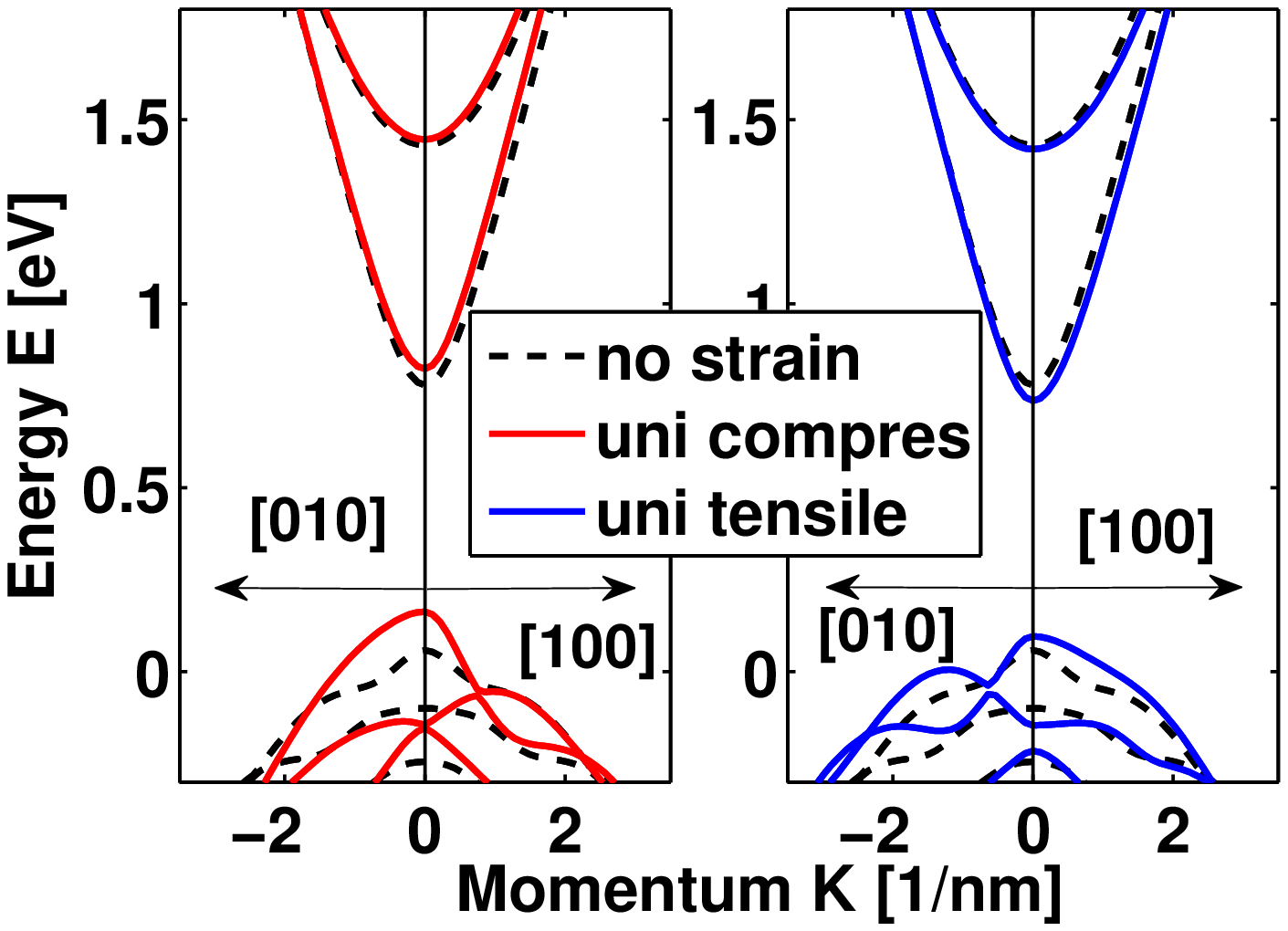}}
\subfigure[]{\includegraphics[width=5.8cm]{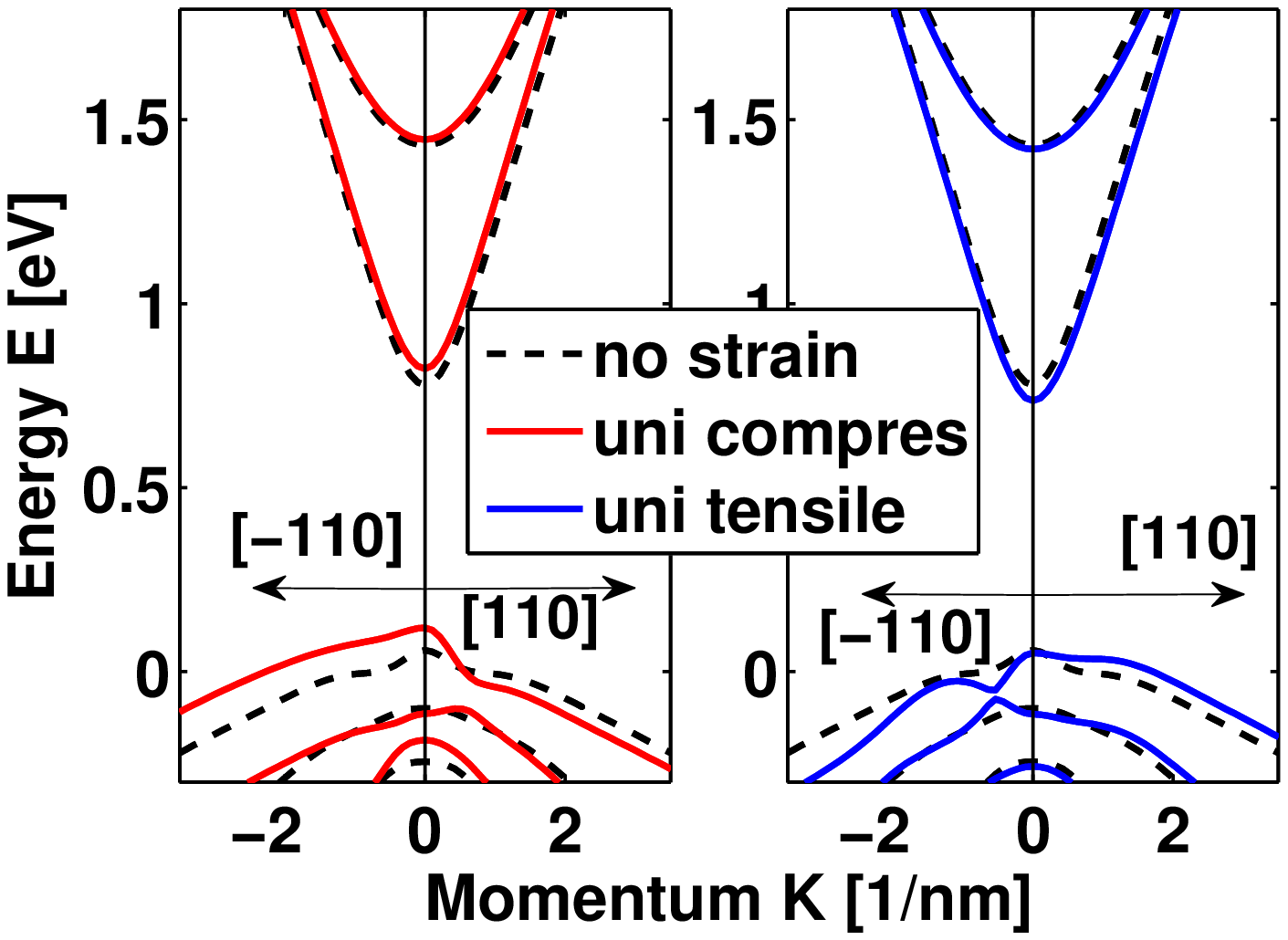}}
\caption{KP calculations of $E$-$k$ diagram for a 3.0nm InAs UTB with 2\% biaxial strain in the (001) plane (a), 2GPa uniaxial stress along [100] direction (b), 2GPa uniaxial stress along [110] direction (c). The confinement direction is along [001] for all cases. The unstrained cases are also shown for comparison.}
\label{fig:ek_strained}
\end{figure*}

The benchmarked $\bf{k}\cdot\bf{p}$ method is then used to study the properties of a 3.0nm thick InAs UTB under various strain conditions.
We consider the most common (001) oriented wafer with a channel direction along [100] and [110]. Note that the direction of the applied uniaxial strain is always aligned with channel (transport) direction. The band structures are plotted in Fig. \ref{fig:ek_strained}, with their band gaps and effective masses at the band edges summarized in Table \ref{tab:gaps_masses}.
Comparing biaxial and uniaxial strain, we find that biaxial strain only slightly changes the band gap or effective masses. The uniaxial strain, instead, significantly modulates the band gap and effective masses. The uniaxial tensile (compressive) strain increases (decreases) transport masses meanwhile decreases (increases) the transverse masses.  Since tunneling probability is an exponential function of the band gap and transport masses \cite{seabaugh2010low}, uniaxial compressive strain is expected to improve TFET $I_{\rm{ON}}$ significantly. In addition, there is a large amount of valence band edge shift under uniaxial compressive strain. Comparing uniaxial compressive strain in the [100] and [110] orientations, we find that [100] strain leads to larger reduction of band gap and transport effective masses and thus is expected to be the most effective in improving $I_{\rm{ON}}$ of TFETs. Thus [100] uniaxial compressive strain is the focus of the remaining study.

\begin{table}
\caption{Band gaps (Eg) and effective masses ($m^*_e$ for electron and $m^*_h$ for hole) extracted from Fig. \ref{fig:ek_strained}.
The percentages in the parentheses are the changes relative to the unstrained cases.}
\label{tab:gaps_masses}       
\begin{tabular}{p{1.3cm}p{1.02cm}p{1.02cm}p{1.02cm}p{1.1cm}p{1.1cm}}
\hline\noalign{\smallskip}
2\% Biaxial  & Eg (eV) & $m^*_e$ [100] & $m^*_e$ [110] & $m^*_h$ [100] & $m^*_h$ [110] \\
\hline\noalign{\smallskip}
No strain           & 0.7206          & 0.0558            &  0.0562          & -0.1106   &  -0.1133 \\
Tensile     & 0.7111 (-1.3\%) & 0.0577 (+3.4\%)   &  0.0581 (+3.4\%) & -0.1272 (+15.0\%)   &  -0.1313 (+15.9\%) \\
Compressive    & 0.7339 (+1.9\%) & 0.0546 (-2.2\%)   &  0.0550 (-2.1\%) & -0.1015 (-8.2\%)   &  -0.1038 (-8.4\%) \\
\hline\noalign{\smallskip}
\hline\noalign{\smallskip}
2GPa Uniaxial [100]  & Eg (eV) & $m^*_e$ [100] & $m^*_e$ [010] & $m^*_h$ [100] & $m^*_h$ [010] \\
\hline\noalign{\smallskip}
No strain       & 0.7206          & 0.0558            &  0.0558          & -0.1106   &  -0.1106 \\
Tensile     & 0.6398 (-11.2\%) & 0.0660 (+18.3\%)   &  0.0477 (-14.5\%) & -0.2917 (+163.7\%)   &  -0.0663 (-40.1\%) \\
Compressive   & 0.6595 (-8.5\%) & 0.0463 (-17.0\%)   &  0.0636 (+14.0\%) & -0.0587 (-46.9\%)   &  -0.2386 (+115.7\%) \\
\hline\noalign{\smallskip}
\hline\noalign{\smallskip}
2GPa Uniaxial [110]  & Eg (eV) & $m^*_e$ [110] & $m^*_e$ [$\bar{1}10$] & $m^*_h$ [110] & $m^*_h$ [$\bar{1}10$] \\
\hline\noalign{\smallskip}
No strain       & 0.7206          & 0.0562            &  0.0562          & -0.1133   &  -0.1133 \\
Tensile     & 0.6852 (-4.9\%) & 0.0620 (+10.3\%)   &  0.0518 (-7.8\%) & -0.3549 (+213.2\%)   &  -0.0704 (-37.9\%) \\
Compressive    & 0.7034 (-2.4\%) & 0.0502 (-10.7\%)   &  0.0598 (+6.4\%) & -0.0637 (-43.8\%)   &  -0.2445 (+115.8\%) \\
\hline\noalign{\smallskip}
\end{tabular}
\end{table}

\section{Globally Strained InAs UTB TFETs}
\begin{figure}[htbp] \centering
{\includegraphics[width=4.35cm]{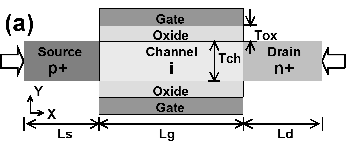}}
{\includegraphics[width=4.35cm]{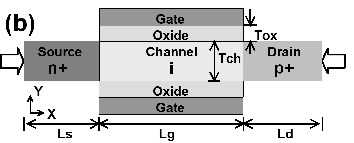}}
\caption{Device structures of UTB nTFET (a) and pTFET (b) under global uniaxial compressive strain. For nTFET, doping density of the source (drain) is $-5\times10^{19}\rm{cm}^{-3}$ ($+5\times10^{18}\rm{cm}^{-3}$), Ls=10nm, Lg=20nm, and Ld=20nm. For pTFET, doping density of the source (drain) is $+1.5\times10^{19}\rm{cm}^{-3}$ ($-2\times10^{19}\rm{cm}^{-3}$), Ls=20nm, Lg=20nm, and Ld=10nm. Channel (InAs UTB) thickness is 3.0nm and equivalent oxide thickness (EOT) is 0.7nm, for both devices. The Z direction is periodic.}
\label{fig:device_global}
\end{figure}

At first, we study InAs UTB TFETs under global (or uniform) strain. The device structures and parameters are illustrated in Fig. \ref{fig:device_global}. We consider 0.3V power supply, i.e., $V_{\rm{DD}}=0.3V$. A lightly doped drain is employed for nTFETs to suppress ambipolar leakage, while a moderately doped source of pTFET is a trade off of Fermi degeneracy and depletion length. Source and drain lengths are different since lower doping density requires a longer depletion region to reach charge neutrality. The confinement and transport orientations are [001] and [100], respectively, with the uniaxial compressive stress applied along [100] as explained above. For high-performance (HP), low operating power (LOP), and low standby power (LSTP) applications, $I_{\rm{OFF}}$ is fixed to $1\times10^{-1}\rm{A/m}$, $1\times10^{-3}\rm{A/m}$, and $1\times10^{-5}\rm{A/m}$, respectively.

\subsection{N-Type TFETs}
\begin{figure}[htbp] \centering
{\includegraphics[width=4.35cm]{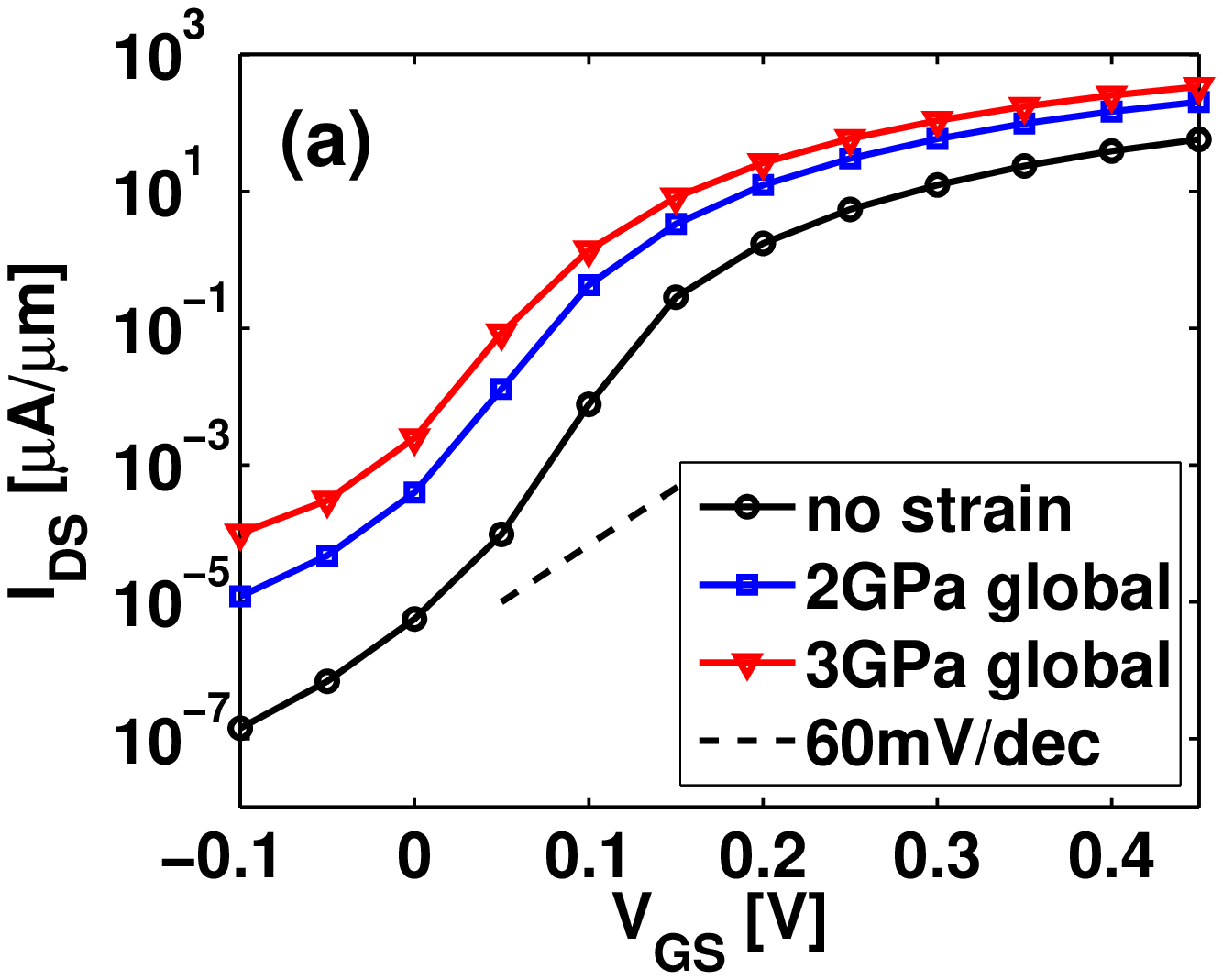}}
{\includegraphics[width=4.35cm]{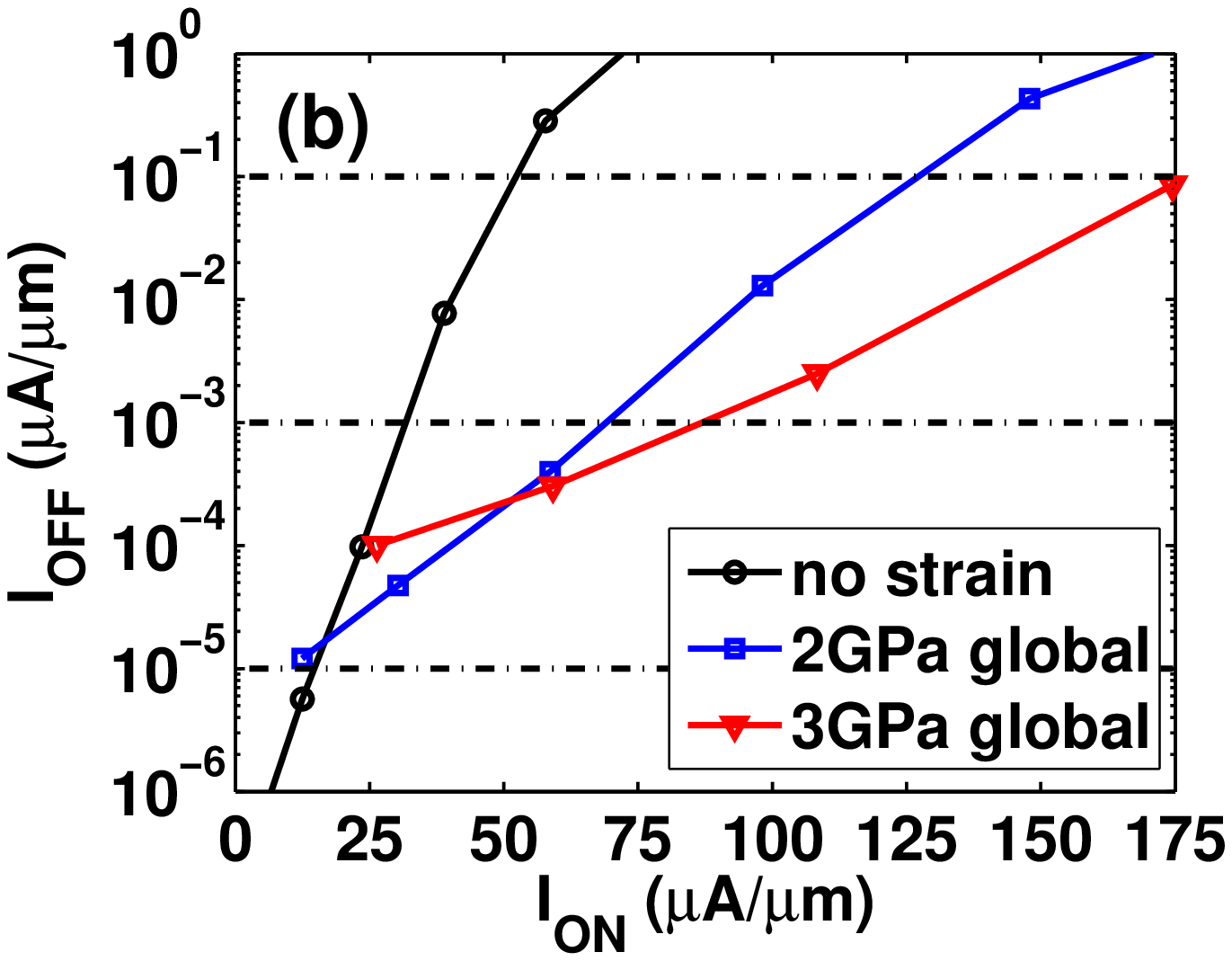}}
{\includegraphics[width=4.35cm]{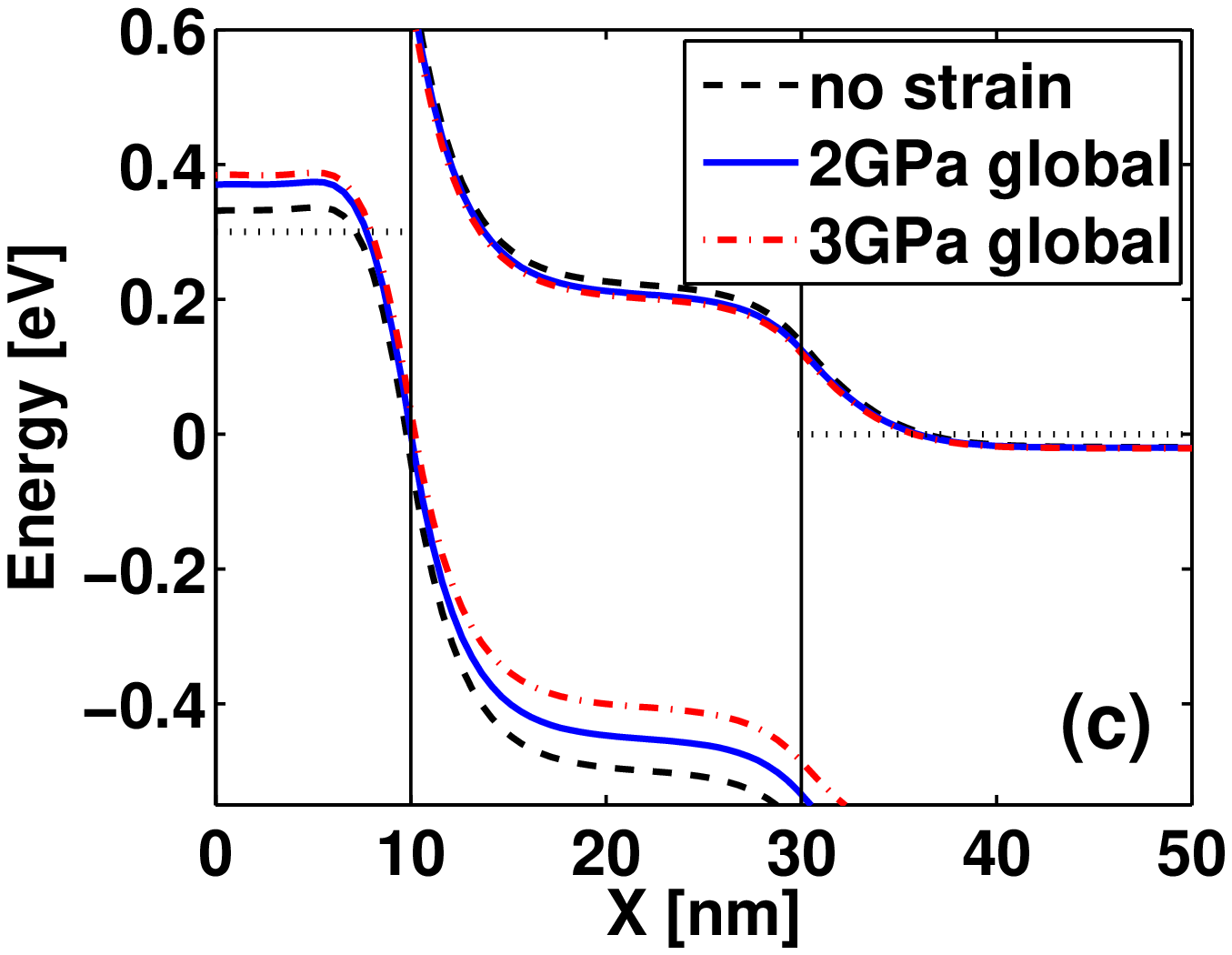}}
{\includegraphics[width=4.35cm]{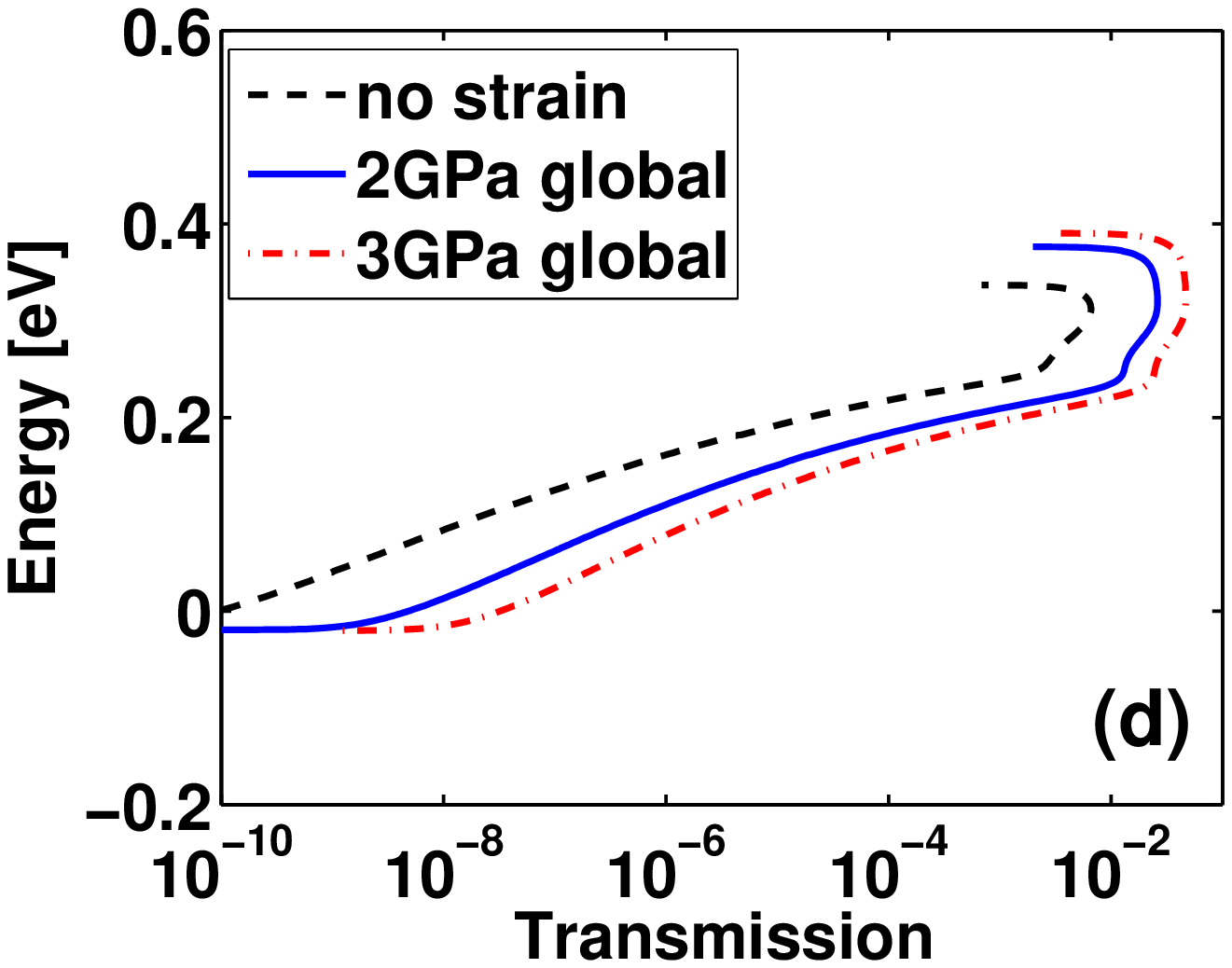}}
\caption{(a) $I_{\rm{DS}}$-$V_{\rm{GS}}$ curve ($V_{\rm{DS}}=0.3V$), (b)$I_{\rm{ON}}$-$I_{\rm{OFF}}$ plot, (c) band diagrams at $V_{\rm{GS}}=0.2V$, and (d) transmissions at $V_{\rm{GS}}=0.2V$ and transverse $k_z=0$, of 2GPa and 3GPa globally strained nTFETs, in comparison with the unstrained case. In (b), HP, LOP, and LSTP applications are marked with dashed lines. In (c), the source and drain Fermi levels are marked with dotted lines.}
\label{fig:n_global}
\end{figure}

Fig. \ref{fig:n_global} compares the nTFETs without strain and with 2GPa/3GPa global uniaxial compressive stress. From (a), it is found that the strain improves the turn-on characteristics, but degrades the SS. As a result, as shown in (b), with fixed $I_{\rm{OFF}}$, $I_{\rm{ON}}$ is improved for HP and LOP applications, but degraded for LSTP application. For LOP application, $I_{\rm{ON}}$ is improved from 32A/m to 69A/m (2GPa) and 87A/m (3GPa). As shown in (c) and (d), the smaller band gap and transport effective masses enabled by the uniaxial compressive strain enhance the transmission, both above and below the channel conduction band edge, leading to not only larger source-to-channel tunnel current but also larger source-to-drain tunnel leakage.

It should be noted that, different from the nanowire case \cite{conzatti2011simulation} where large source Fermi degeneracy is created by strain, the source Fermi degeneracy for the strained cases here is only about 0.05eV higher than the unstrained case so it does not appreciably degrade the SS. This is because although the hole transport effective mass is reduced by strain the hole transverse effective mass is increased, as shown in Table \ref{tab:gaps_masses}, and therefore the source DOS is only slightly reduced.

\subsection{P-Type TFETs}
\begin{figure}[htbp] \centering
{\includegraphics[width=4.35cm]{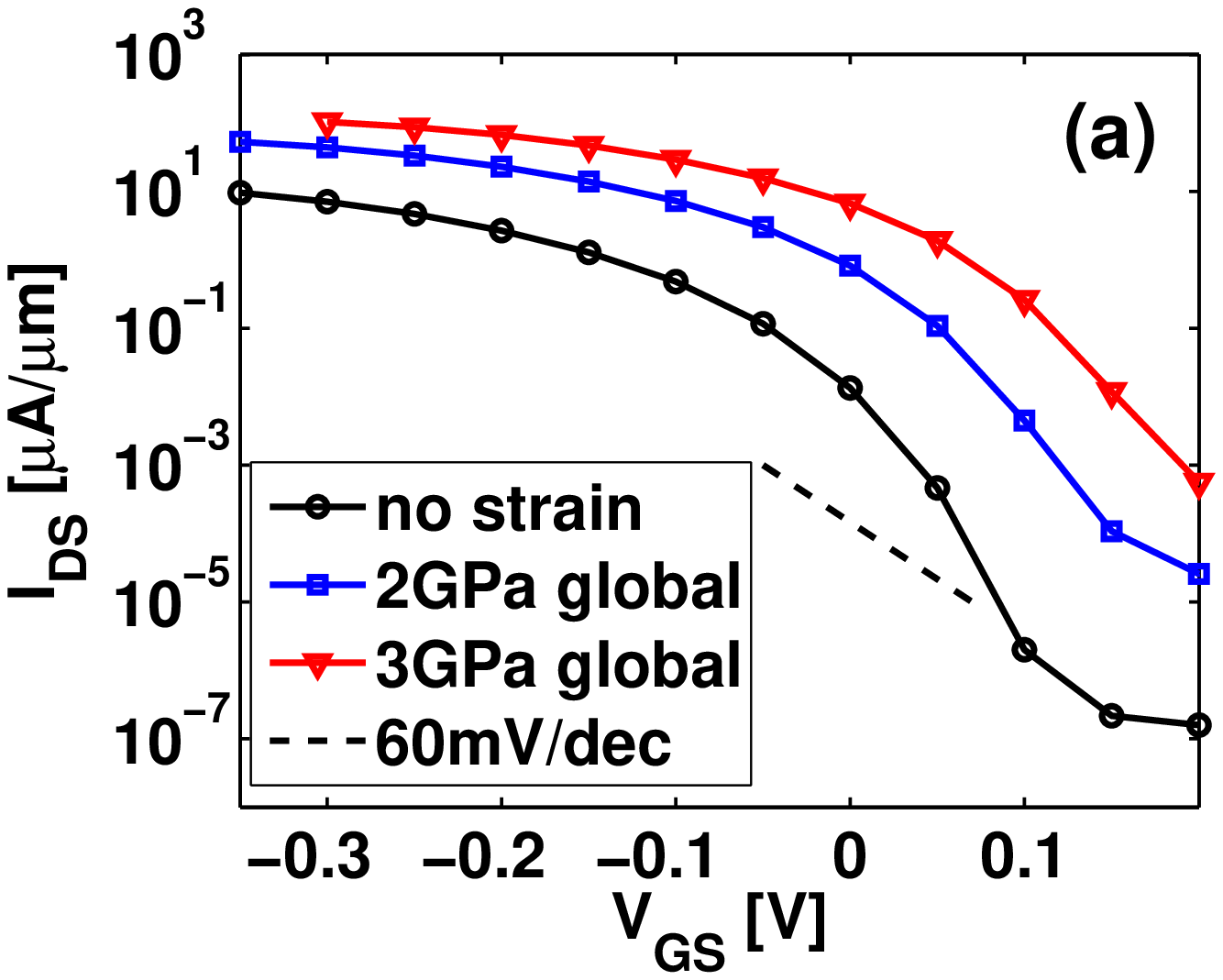}}
{\includegraphics[width=4.35cm]{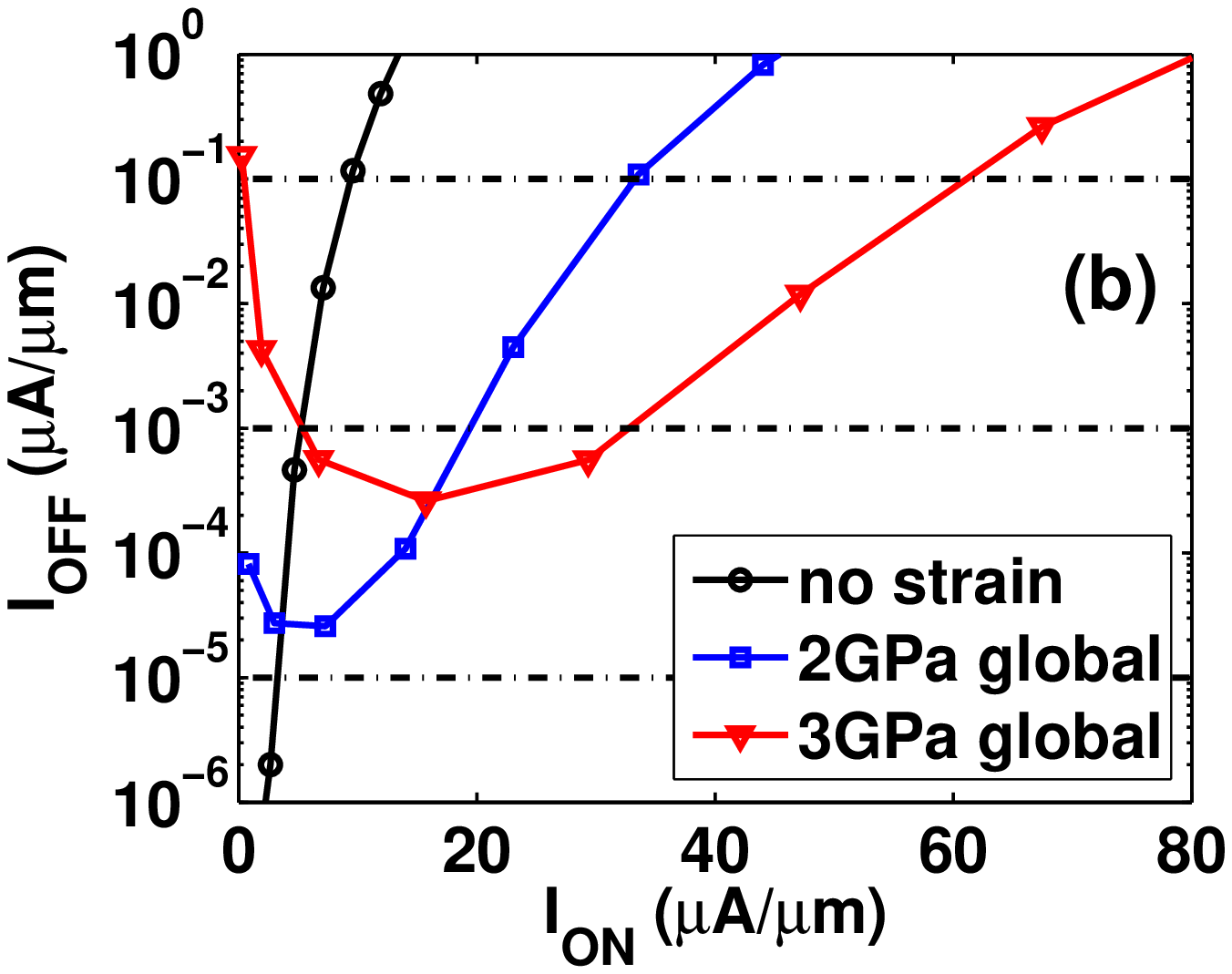}}
{\includegraphics[width=4.35cm]{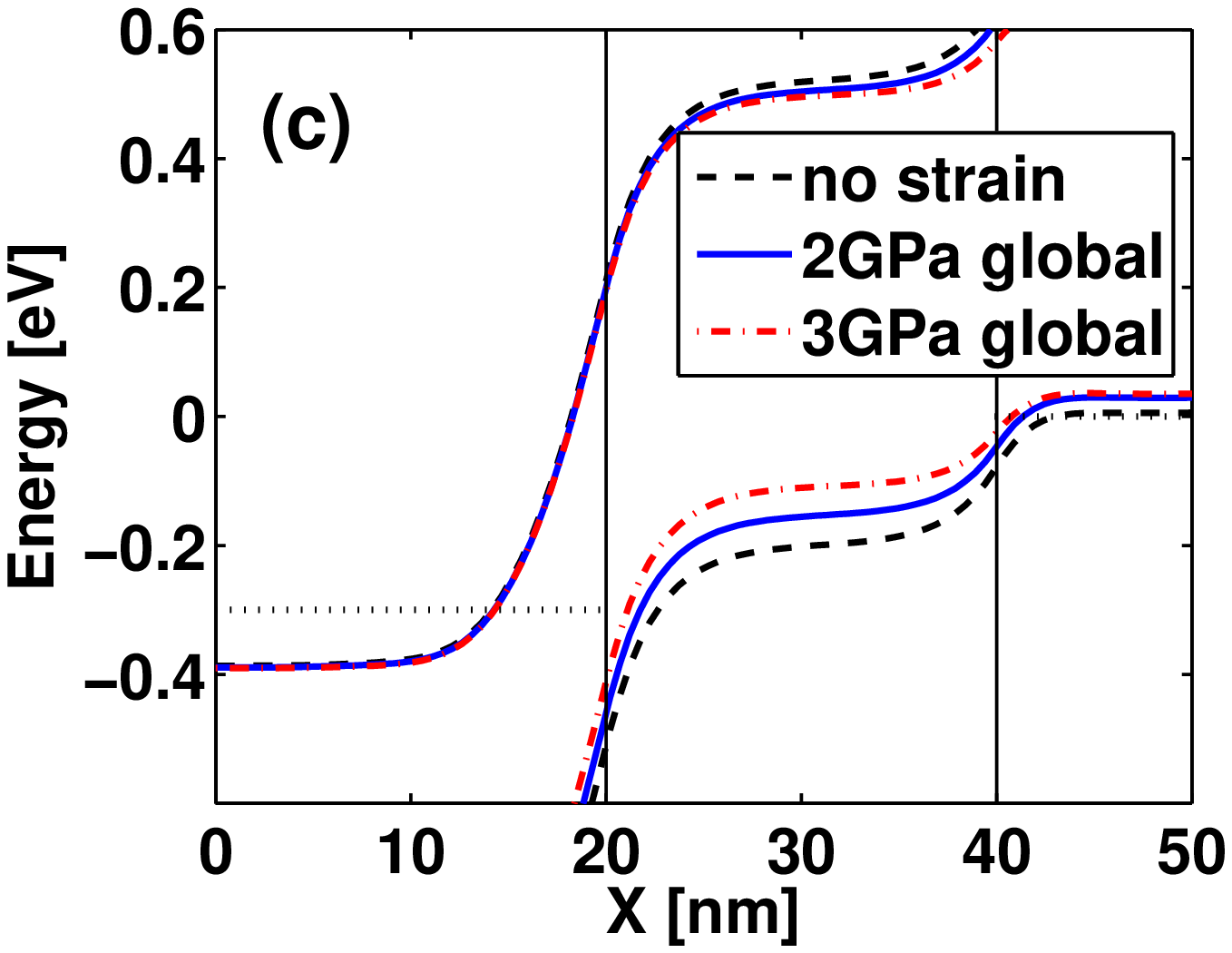}}
{\includegraphics[width=4.35cm]{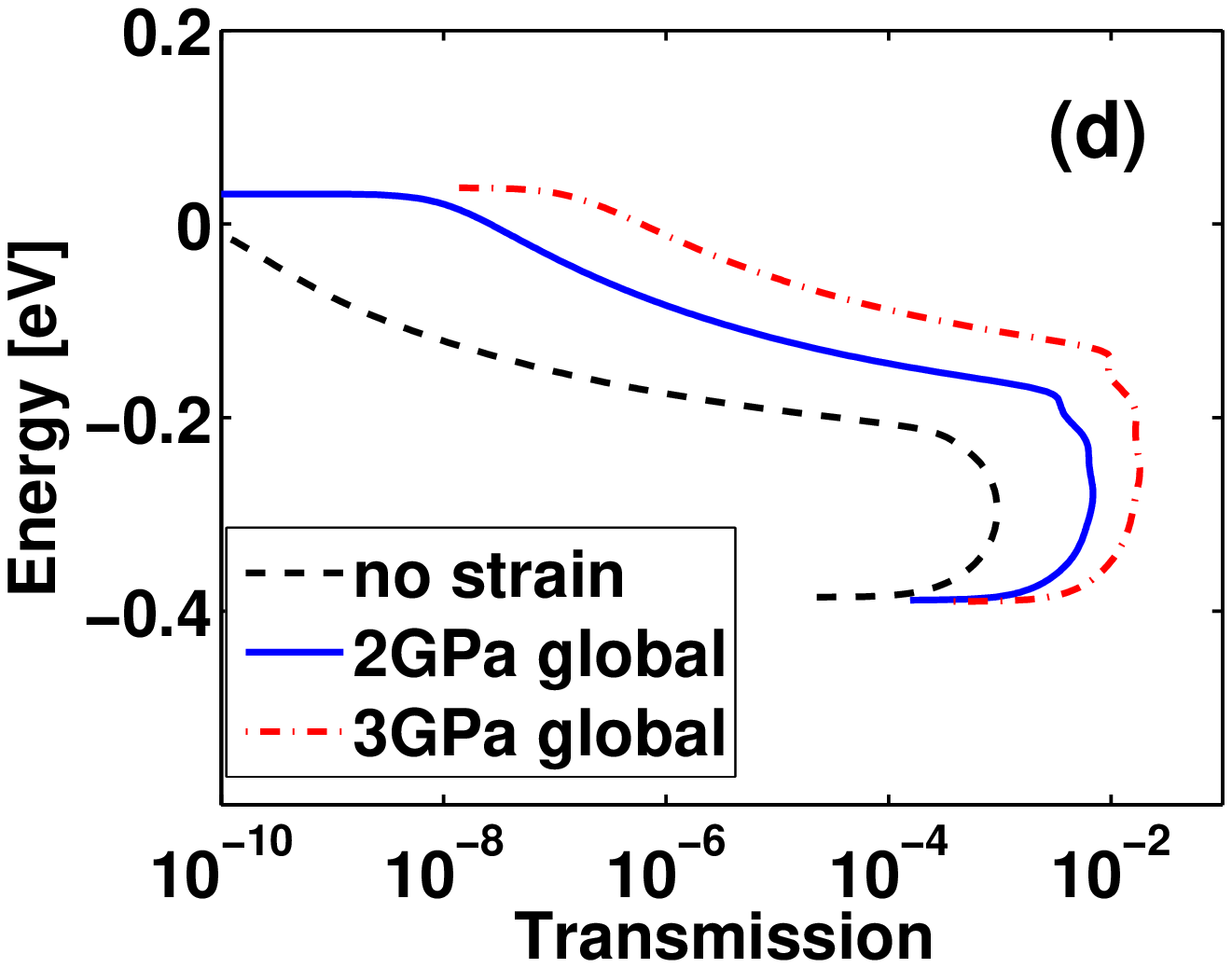}}
\caption{(a) $I_{\rm{DS}}$-$V_{\rm{GS}}$ curve ($V_{\rm{DS}}=-0.3V$), (b)$I_{\rm{ON}}$-$I_{\rm{OFF}}$ plot, (c) band diagrams at $V_{\rm{GS}}=-0.1V$, and (d) transmissions at $V_{\rm{GS}}=-0.1V$ and transverse $k_z=0$, of 2GPa and 3GPa globally strained pTFETs, in comparison with the unstrained case. In (b), HP, LOP, and LSTP applications are marked with dashed lines. In (c), the source and drain Fermi levels are marked with dotted lines.}
\label{fig:p_global}
\end{figure}

Fig. \ref{fig:p_global} compares the pTFETs without strain and with 2GPa/3GPa global uniaxial compressive stress. Similar to the n-type cases, the strain improves $I_{\rm{ON}}$ of pTFETs for both HP and LOP applications, but degrades $I_{\rm{ON}}$ for LSTP application. For LOP application, $I_{\rm{ON}}$ is improved from 5A/m to 19A/m (2GPa) and 33A/m (3GPa). It is seen that the source Fermi degeneracy does not change with the strain, as the density of states (DOS) of the conduction band does not change much with the strain (the electron transport mass decreases but the electron transverse mass increases, as shown in Table \ref{tab:gaps_masses}).

\section{Locally Strained InAs UTB TFETs}
\begin{figure}[htbp] \centering
{\includegraphics[width=4.35cm]{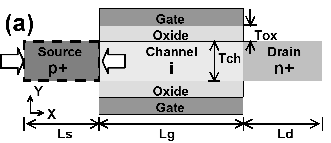}}
{\includegraphics[width=4.35cm]{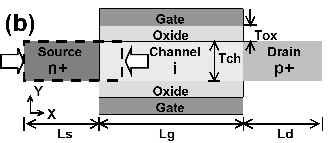}}
\caption{Device structures of UTB nTFET (a) and pTFET (b) under local uniaxial compressive strain. The device parameters are the same as those in Fig. \ref{fig:device_global}. Note the difference of the strained areas.}
\label{fig:device_local}
\end{figure}

Local (or non-uniform) strain can be used to overcome the drawbacks of globally strained TFETs, i.e., large source-to-drain and ambipolar leakage, and at the same time to retain the advantages, i.e., large source-to-channel transmission.
As illustrated in Fig. \ref{fig:device_local}, the local uniaxial compressive strain is applied only to the source portion of nTFET and pTFET. For pTFET, the locally strained region is slightly extended into the channel to induce further improvement, as will be explained in the following. The feasibility of fabricating such locally strained TFETs has been discussed in \cite{Boucart2009lateral}.

\subsection{N-Type TFETs}
\begin{figure}[htbp] \centering
{\includegraphics[width=4.35cm]{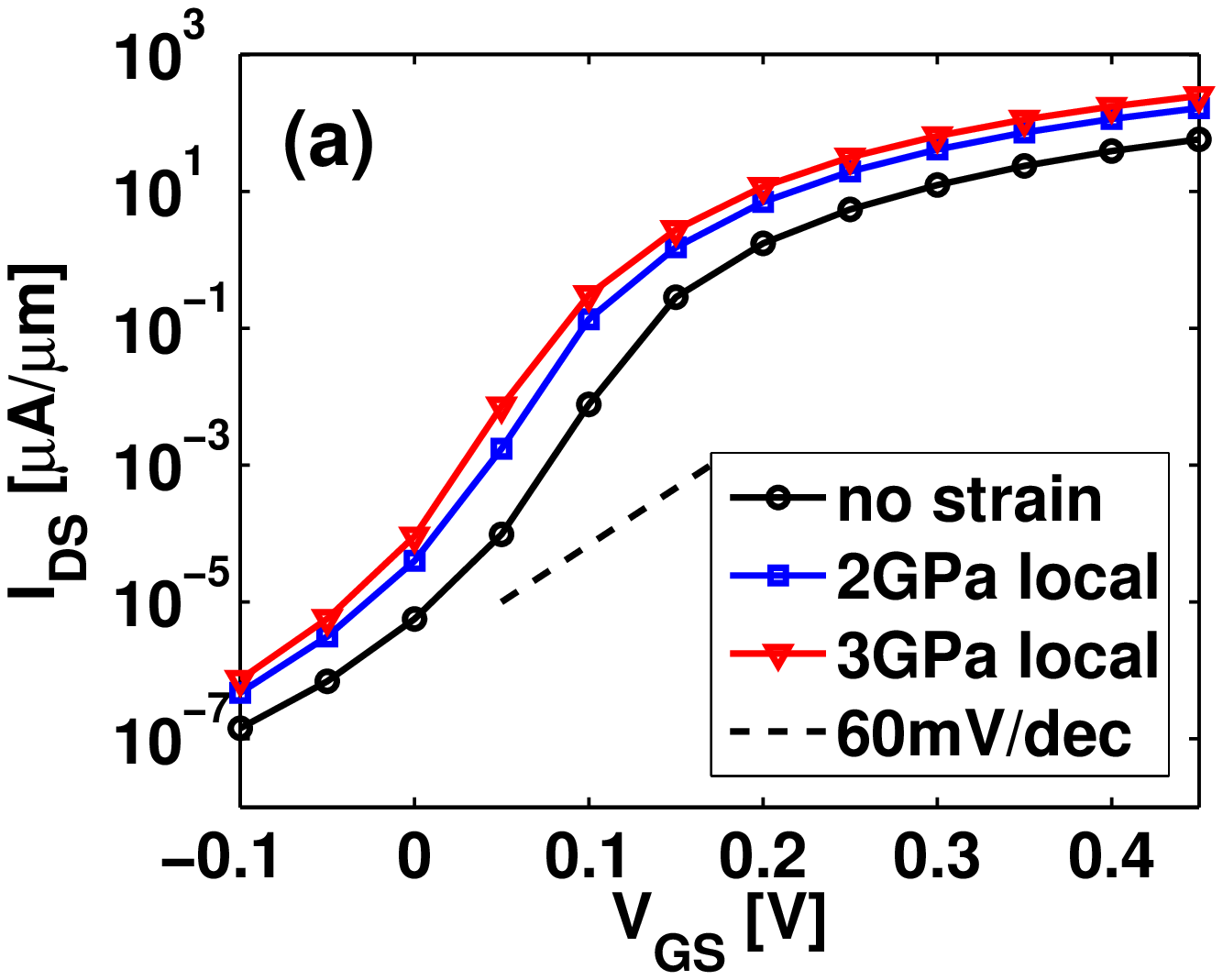}}
{\includegraphics[width=4.35cm]{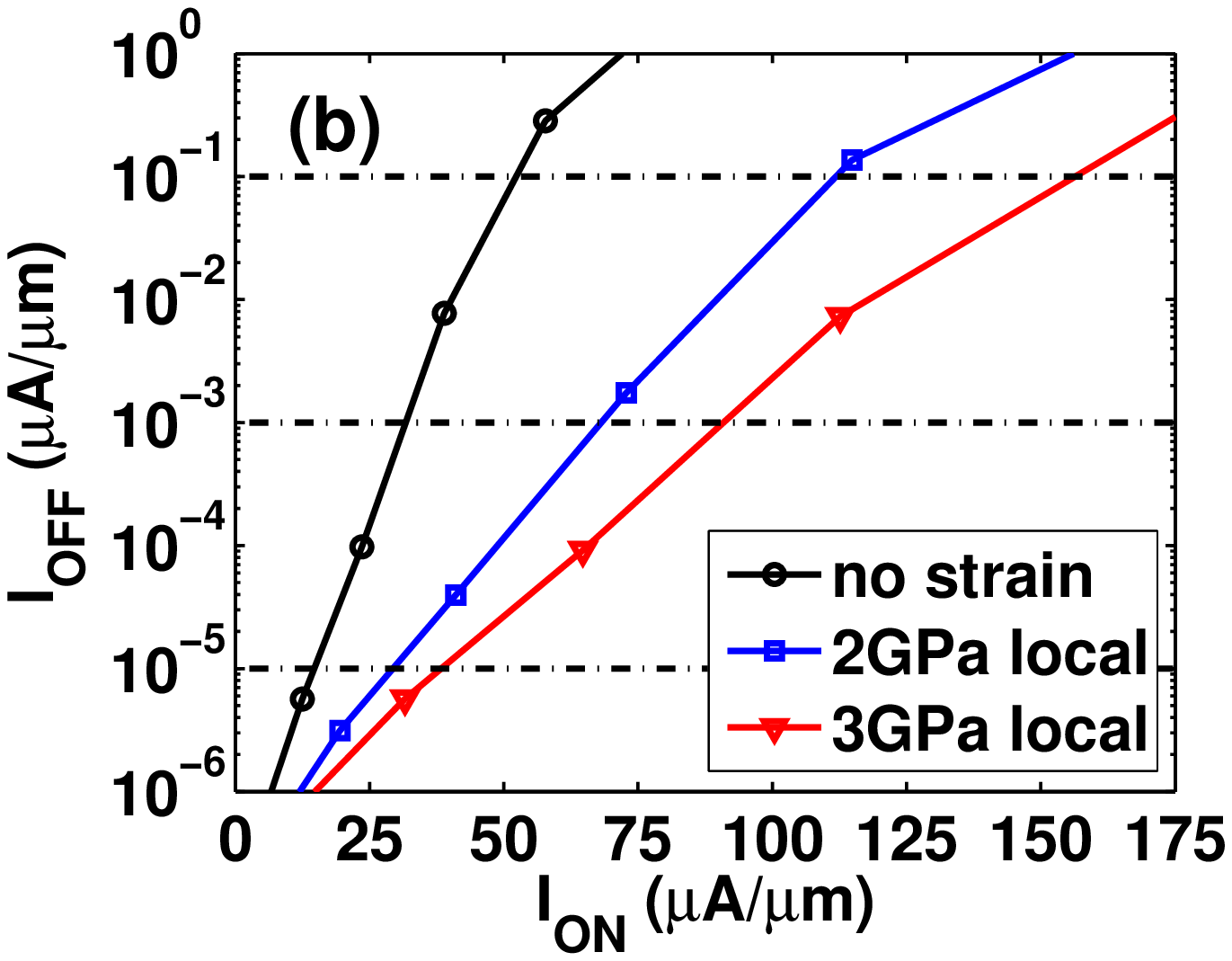}}
{\includegraphics[width=4.35cm]{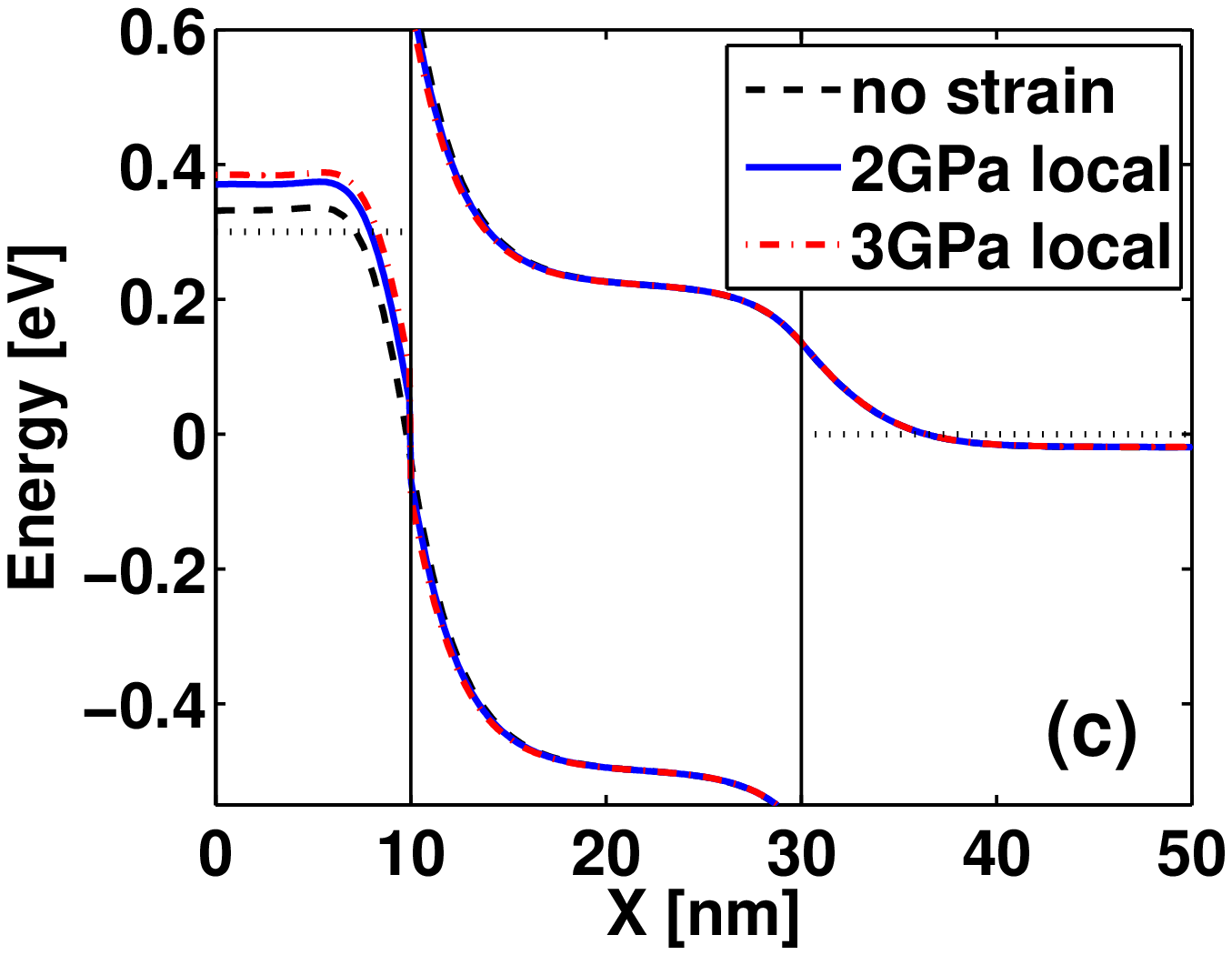}}
{\includegraphics[width=4.35cm]{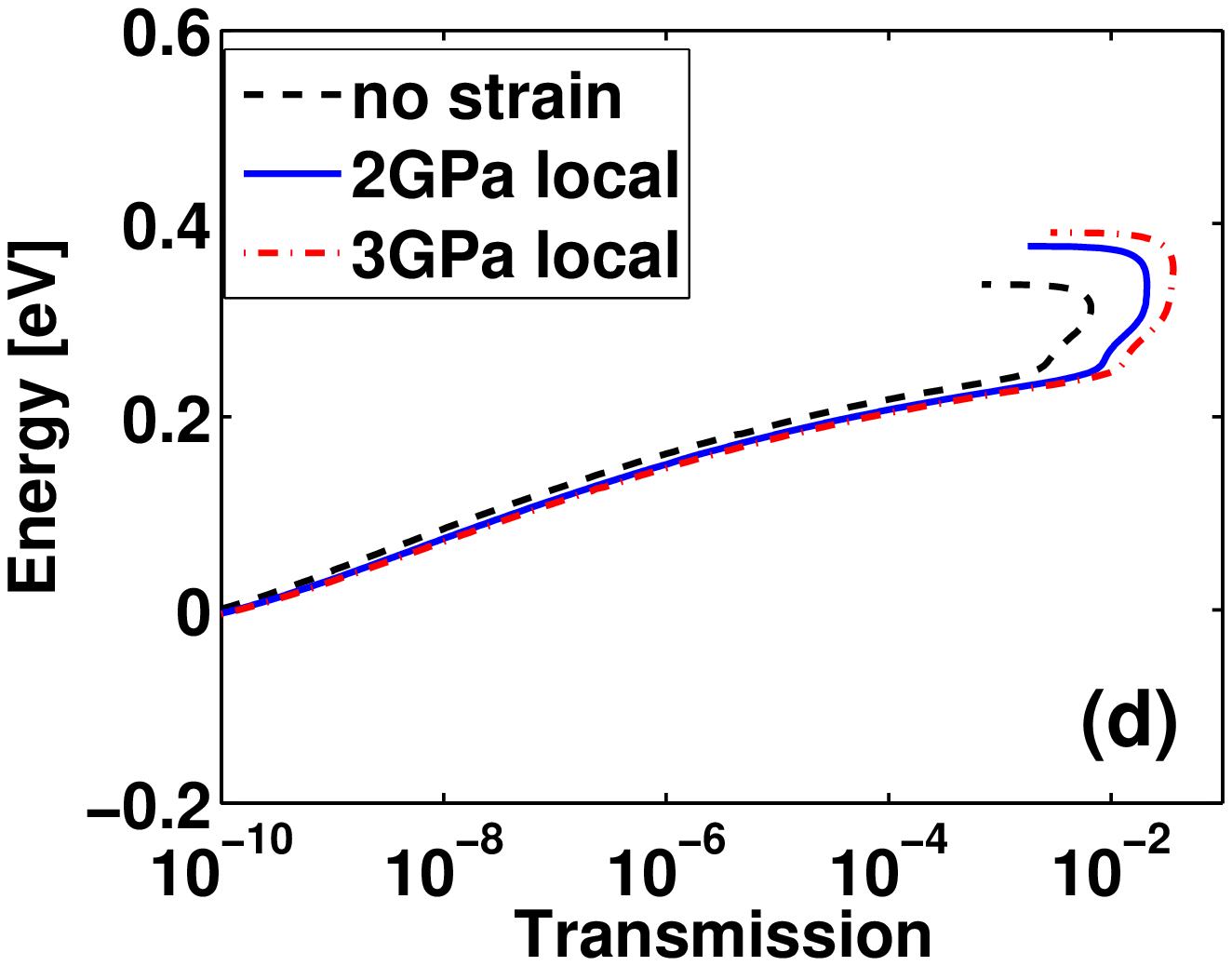}}
\caption{The same plots as Fig. \ref{fig:n_global} but for locally strained nTFETs.}
\label{fig:n_local}
\end{figure}

Fig. \ref{fig:n_local} compares nTFETs without strain and with 2GPa/3GPa local uniaxial compressive strain in the source. It is found that the strain improves $I_{\rm{ON}}$ for all HP, LOP, and LSTP applications. For LOP application, $I_{\rm{ON}}$ is improved from 32A/m to 68A/m (2GPa) and 91A/m (3GPa), which is similar to the global strain case. For LSTP application, $I_{\rm{ON}}$ is improved from 15A/m to 29A/m (2GPa) and 38A/m (3GPa). As shown in (c), the band gap and transport effective mass decrease in the source, but remain unchanged in the channel. As seen in (d), the source-to-channel tunnel probability still increases, as in the global strain case; but the source-to-drain leakage and ambipolar leakage are not affected, in contrast to the global strain case.

\subsection{P-Type TFETs}
\begin{figure}[htbp] \centering
{\includegraphics[width=4.35cm]{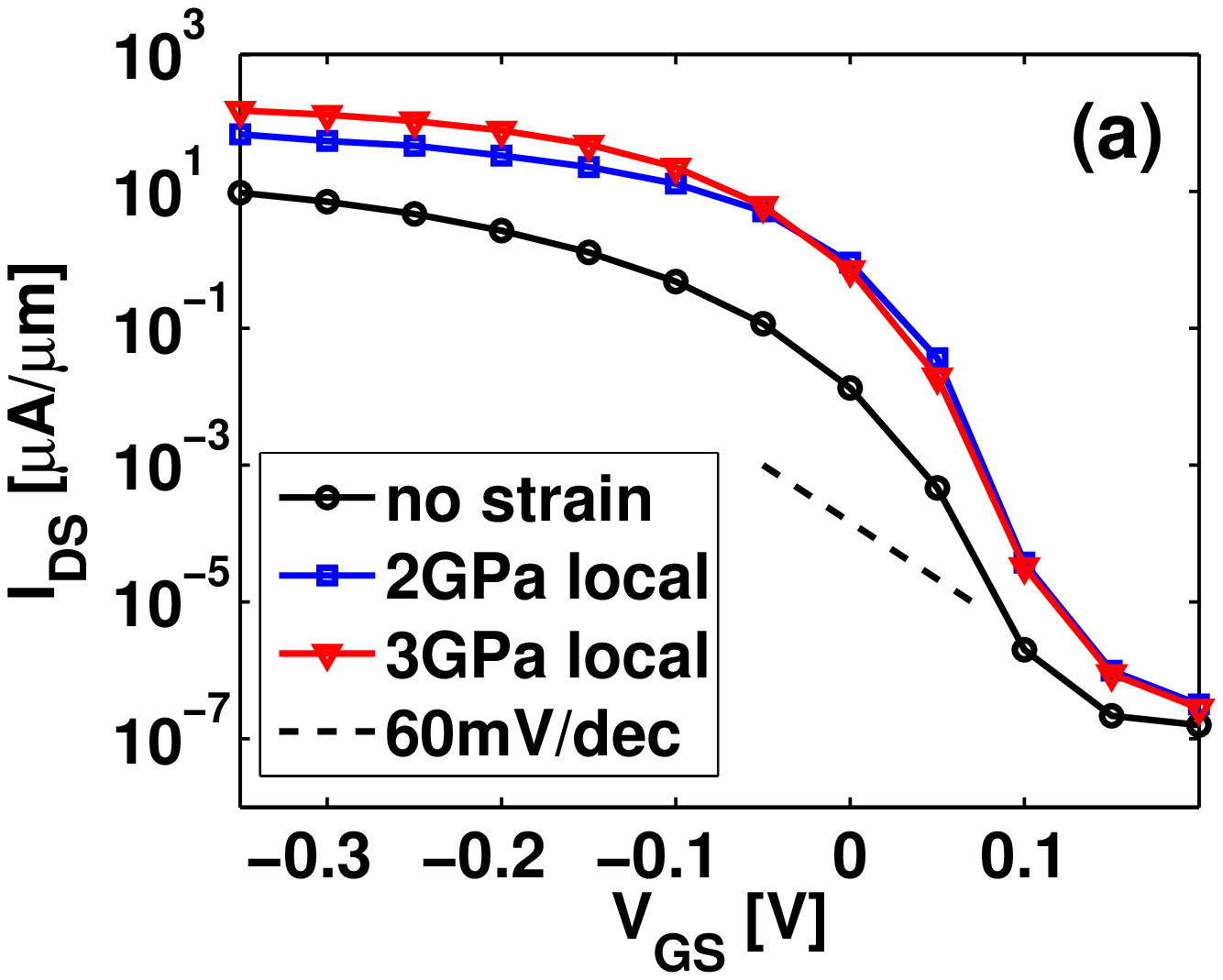}}
{\includegraphics[width=4.35cm]{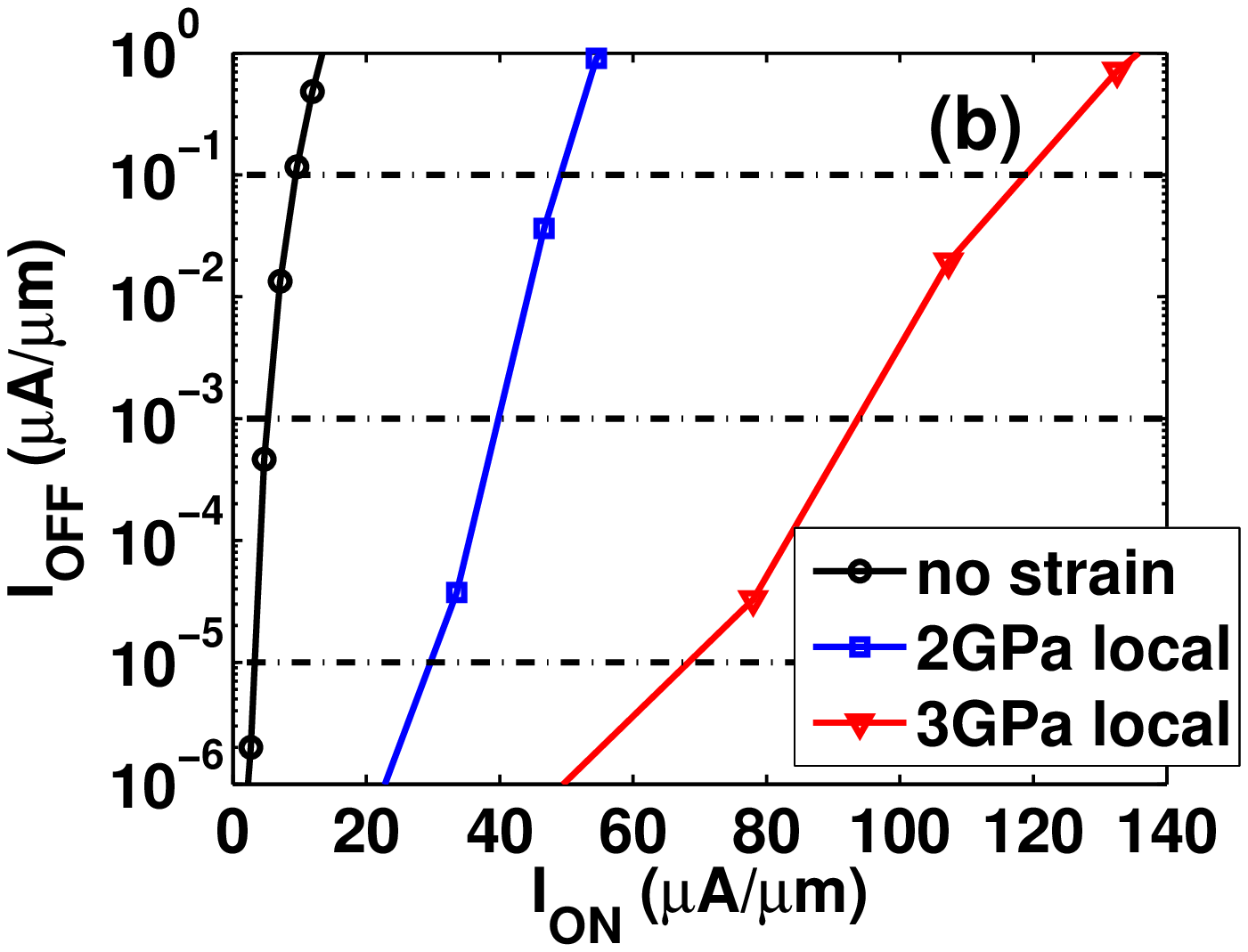}}
{\includegraphics[width=4.35cm]{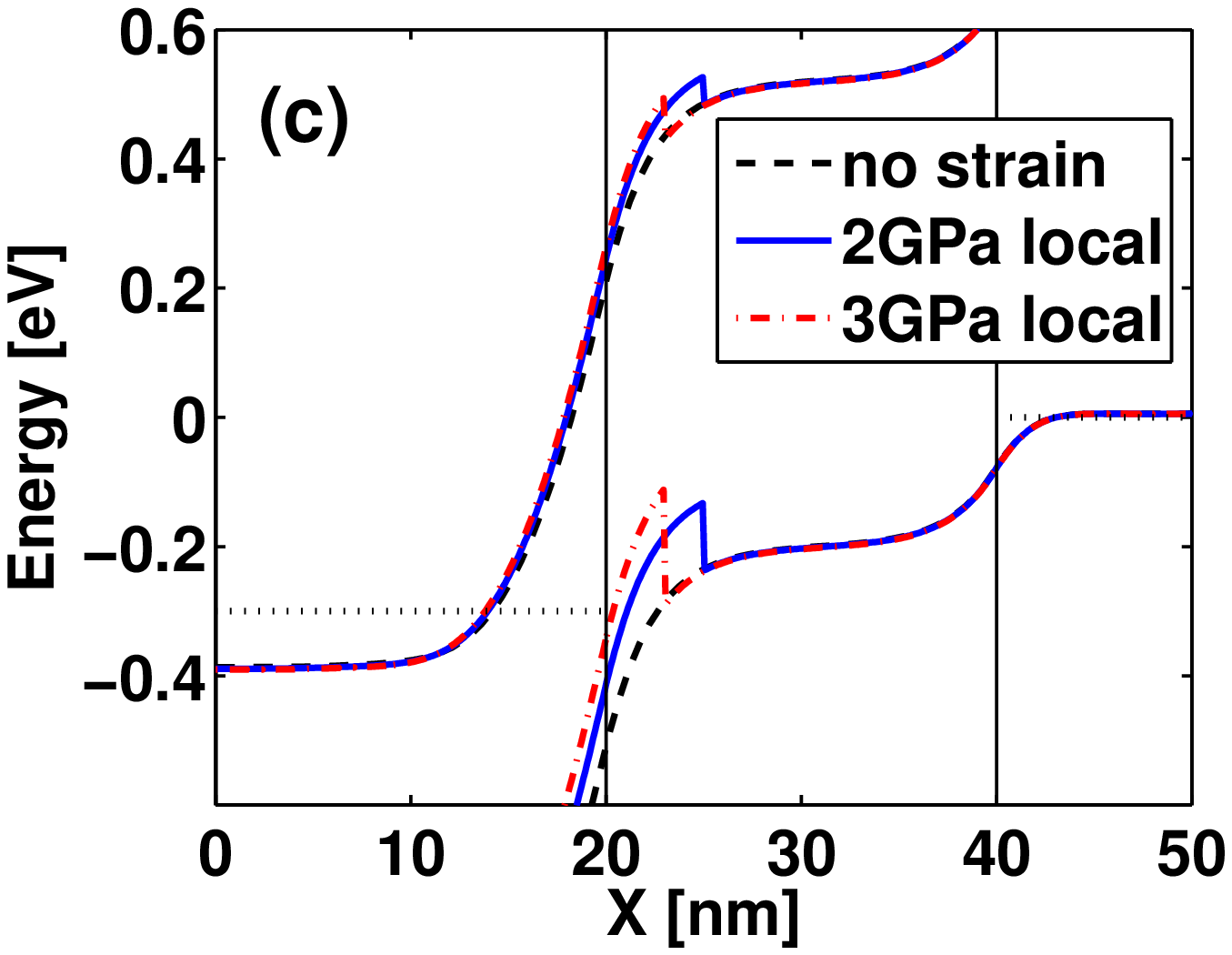}}
{\includegraphics[width=4.35cm]{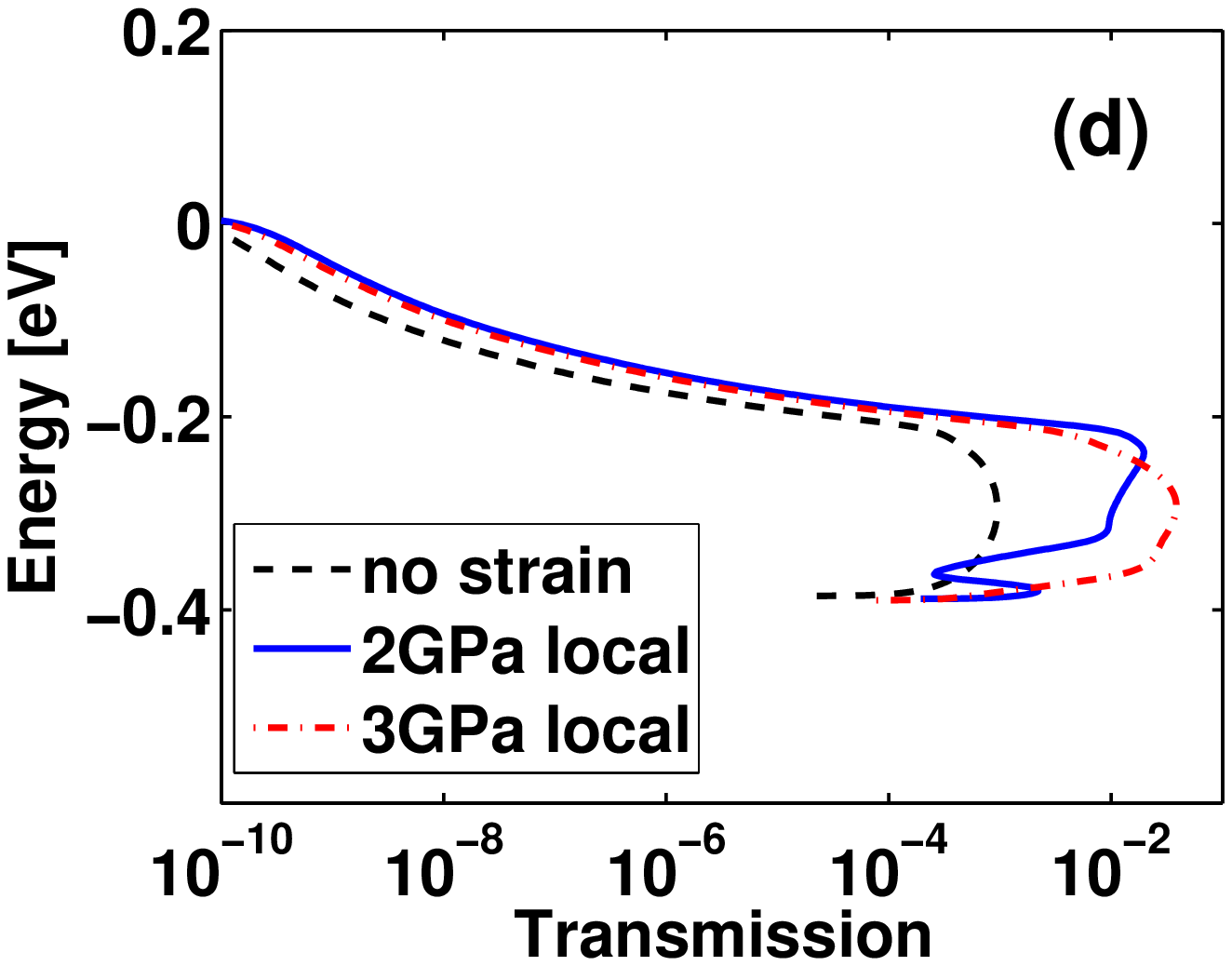}}
\caption{The same plots as Fig. \ref{fig:p_global} but for locally strained pTFETs.}
\label{fig:p_local}
\end{figure}

Fig. \ref{fig:p_local} compares the pTFETs without strain and with 2GPa/3GPa local uniaxial compressive strain in the source and in the 5nm/3nm extension region. It is found that the strain improves $I_{\rm{ON}}$ for all HP, LOP, and LSTP applications. For LOP application, $I_{\rm{ON}}$ is improved from 5A/m to 40A/m (2GPa) and 94A/m (3GPa), larger than the improvement enabled by global strain. For LSTP application, $I_{\rm{ON}}$ is improved from 3A/m to 30A/m (2GPa) and 68A/m (3GPa). Similar to the locally strained nTFET cases, the source-to-channel tunnel probability increases while the source-to-drain leakage and ambipolar leakage are not infuenced. In addition, the source-to-channel tunnel barrier thickness is further reduced by the valence band discontinuity (between strained and unstrained parts), similar to the channel heterojunction design \cite{Long2016}. For nTFETs, unfortunately, we could not employ the same design, i.e., extending the locally strained area into the channel and make use of the conduction band discontinuity. This is because the conduction band edge moves upward instead of downward under uniaxial compressive strain. Thus, the improvement of p-type cases is more significant than the n-type cases.

\begin{figure}[htbp] \centering
{\includegraphics[width=4.35cm]{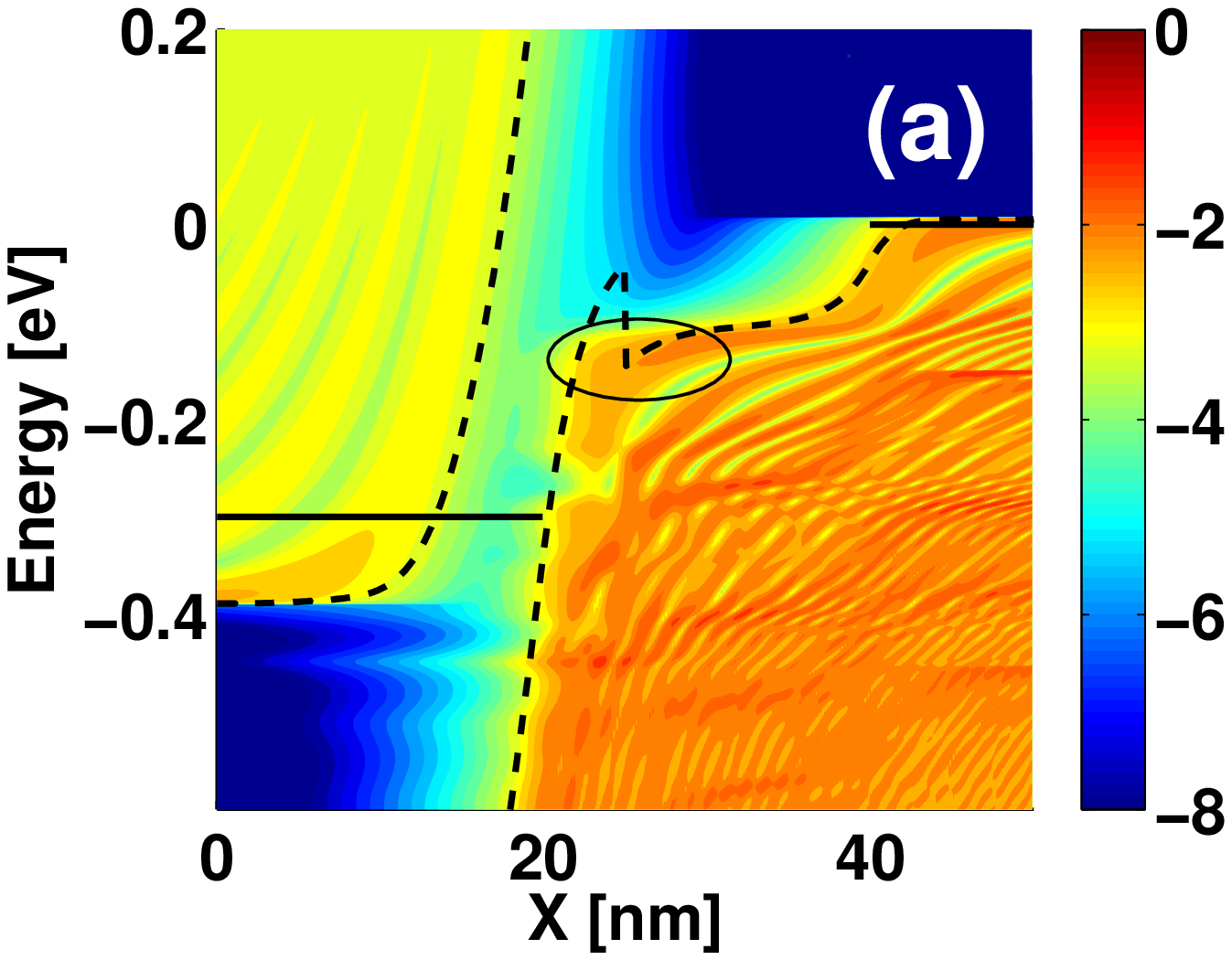}}
{\includegraphics[width=4.35cm]{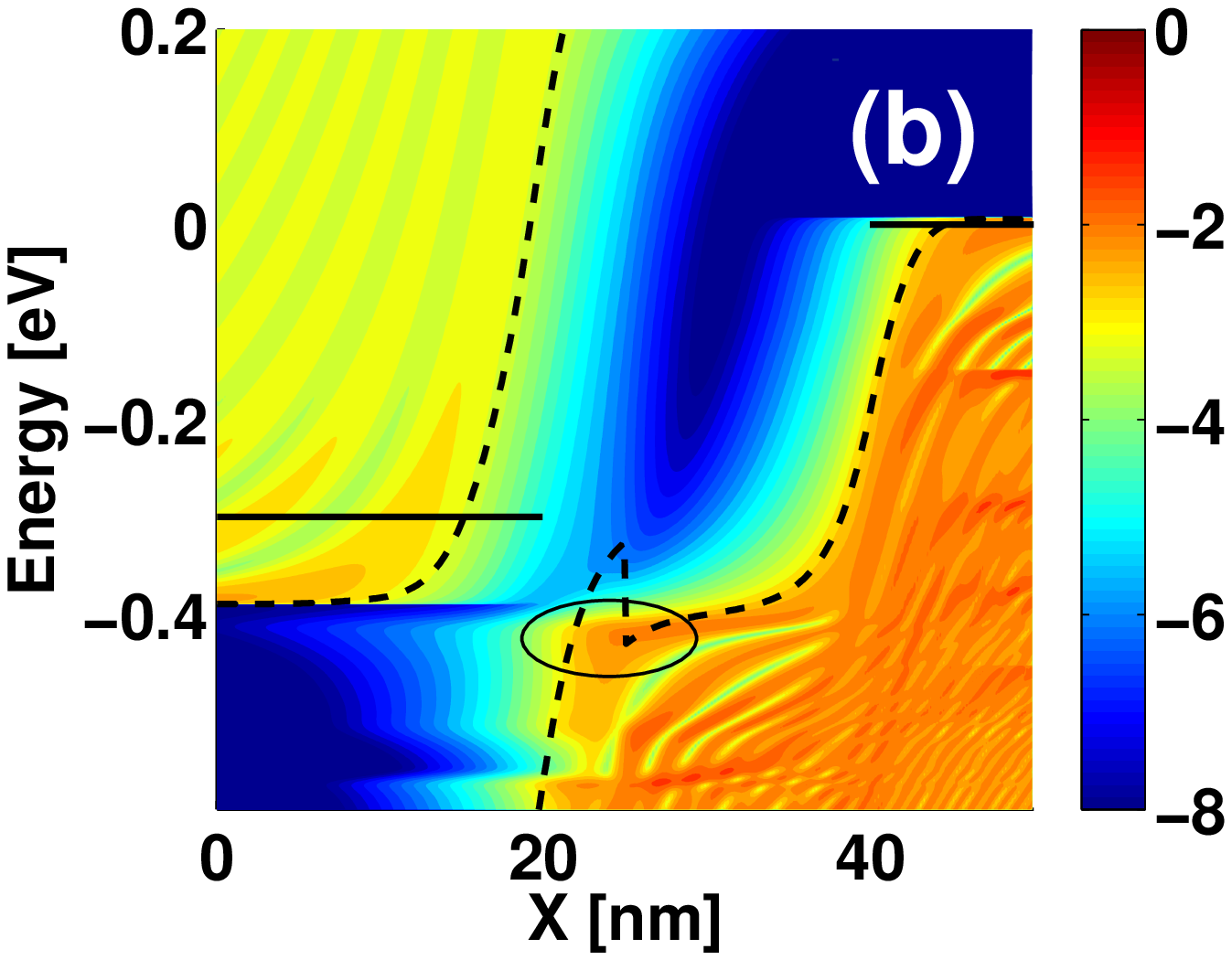}}
\caption{LDOS (in logarithmic scale) of the locally strained pTFET with 2GPa stress, around ON state with $V_{\rm{GS}}=-0.2V$ (a) and around OFF state with $V_{\rm{GS}}=0.1V$ (b). Band diagrams (dashed lines) and contact Fermi levels (solid lines) are superimposed. The notch states are highlighted with circles.}
\label{fig:ldos}
\end{figure}

Fig. \ref{fig:ldos} (a) and (b) depict the local density of states (LDOS) of the locally strained pTFET with 2GPa stress, around ON and OFF state, respectively. At ON state, the band discontinuity forms a triangular quantum well, creating a resonant notch state aligned with the channel valence band edge, enhancing the transmission. At OFF state, the notch state lies outside the quantum well, reducing the thermal emission induced leakage. When the stress is increased from 2GPa to 3GPa, the band discontinuity increases, and the extension region needs to be reduced from 5nm to 3nm so as to avoid forming the notch state inside the quantum well. The phenomenon has also been observed in channel heterojunction TFETs \cite{Long2016}.

It should be mentioned that the local strain is assumed to be abrupt in this study. In practice, there should be a transition region between the strained and the relaxed portions, which will smooth out the abrupt band profile and may change the simulation results quantitatively. We have not performed such kind of simulation since the actual strain profile depends on specific strain implementation. Note that studies in \cite{Boucart2009lateral,Conzatti2013} assume Gaussian strain profiles.

\begin{figure}[htbp] \centering
{\includegraphics[width=4.35cm]{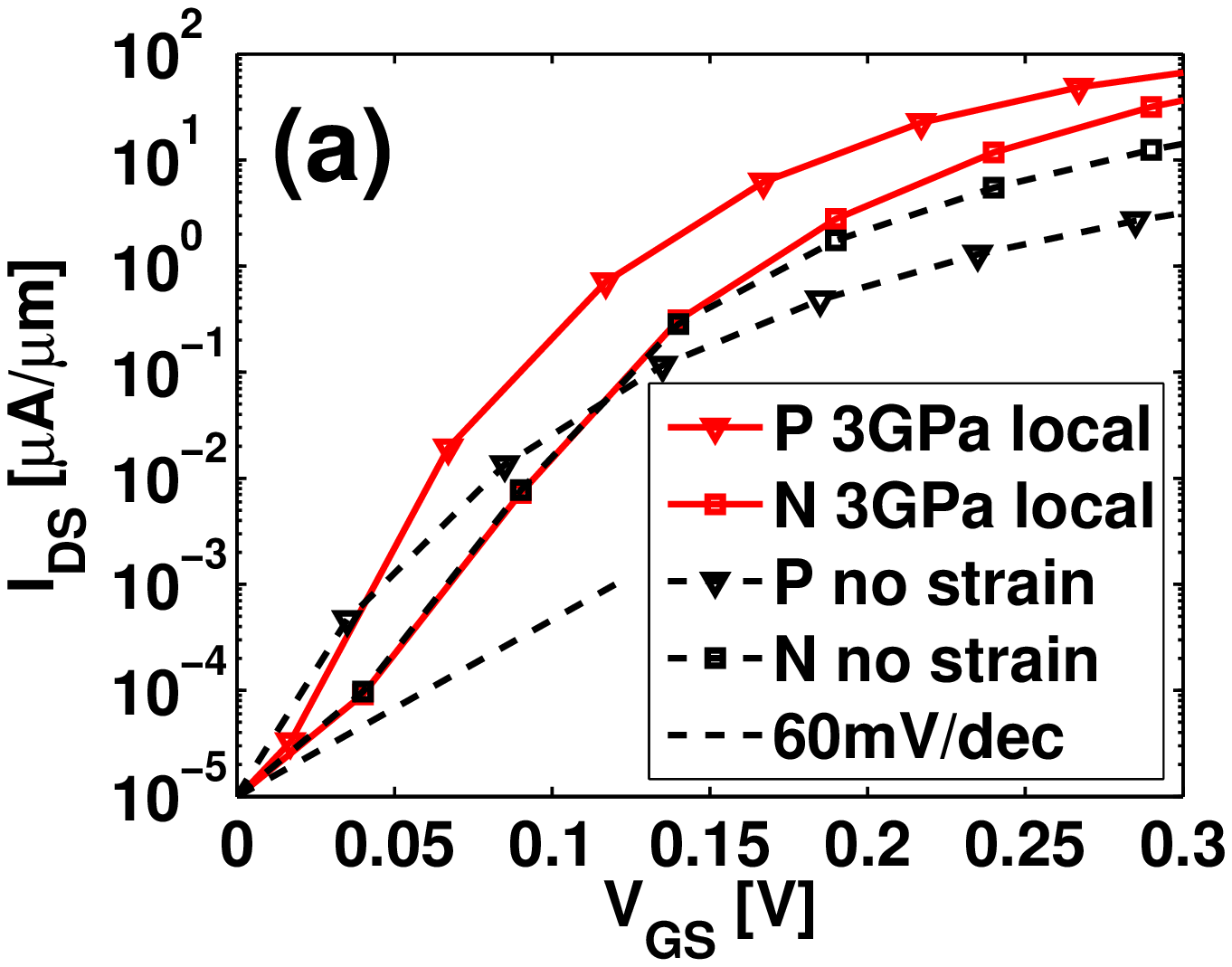}}
{\includegraphics[width=4.35cm]{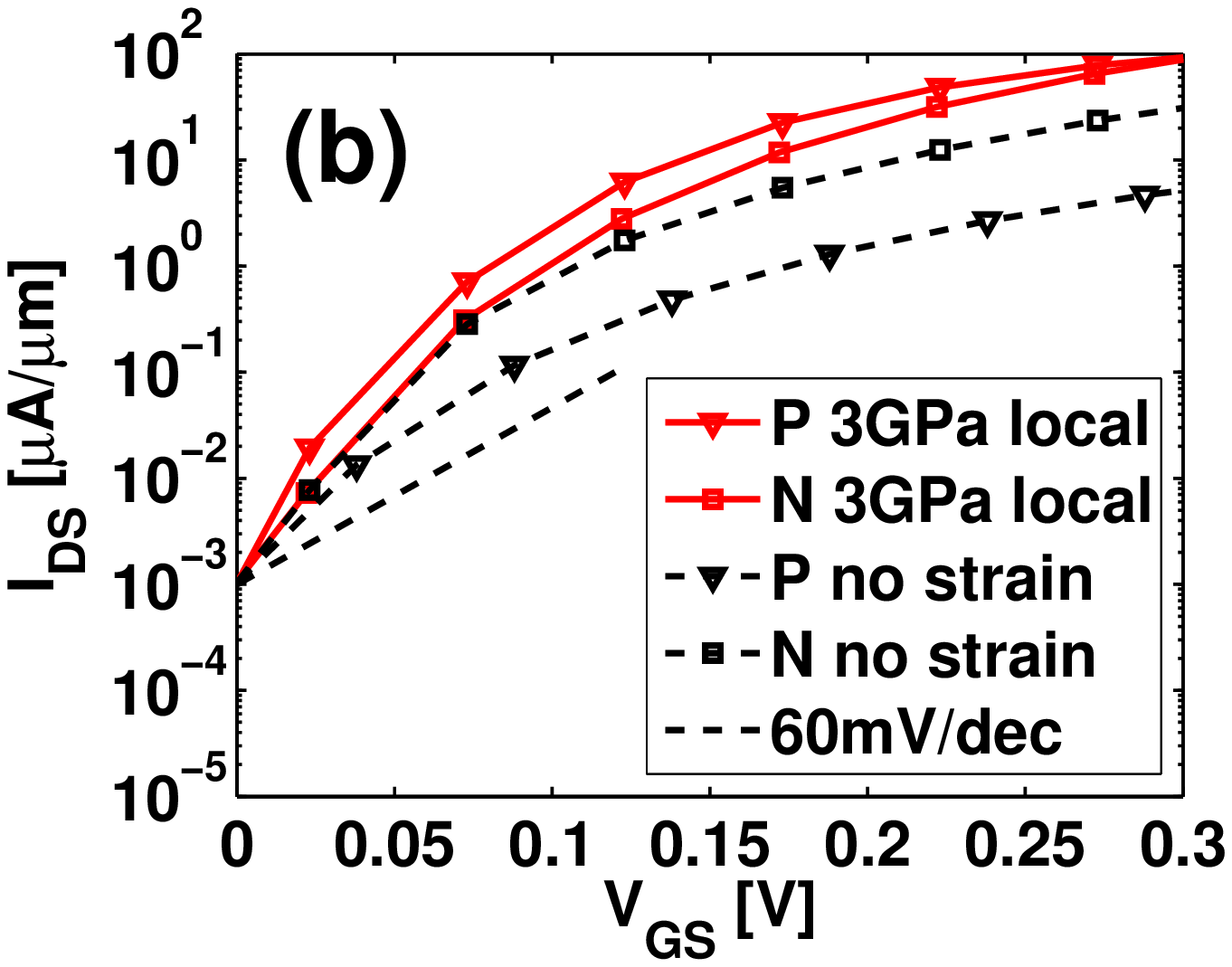}}
\caption{$I_{\rm{DS}}$-$V_{\rm{GS}}$ characteristics ($V_{\rm{DD}}=0.3V$) of P and N type InAs UTB TFETs before and after 3GPa local uniaxial compressive stress, at two $I_{\rm{OFF}}$ targets. (a) $I_{\rm{OFF}}=1\times10^{-5}\rm{A/m}$, and (b) $I_{\rm{OFF}}=1\times10^{-3}\rm{A/m}$.}
\label{fig:p_vs_n}
\end{figure}

\section{Conclusion}
Quantum ballistic transport simulations employing eight-band $\mathbf{k}\cdot\mathbf{p}$ Hamiltonian are carried out to study strain effects on UTB III-V TFETs. It is found, that for an InAs UTB confined in the [001] orientation, uniaxial compressive stress in the [100] orientation significantly reduces the band gap and transport masses, meanwhile increases the transverse masses, and thus can be employed to improve InAs UTB TFETs. Larger improvements can be obtained by applying the strain locally in the source. As compared in Fig. \ref{fig:p_vs_n}, for an n-type (p-type) InAs UTB TFET with 20nm channel length and 3nm body thickness, at $V_{\rm{DD}}=0.3V$ and $I_{\rm{OFF}}=1\times10^{-5}\rm{A/m}$, the strain improves $I_{\rm{ON}}$ from 15A/m (3A/m) to 38A/m (68A/m). While at $V_{\rm{DD}}=0.3V$ and $I_{\rm{OFF}}=1\times10^{-3}\rm{A/m}$, it improves $I_{\rm{ON}}$ from 32A/m (5A/m) to 91A/m (94A/m). The strain not only improves $I_{\rm{ON}}$ of both n-type and p-type devices but also narrows the performance gap between them. Therefore we think strain engineering is a promising performance booster for complementary circuits based on UTB III-V TFETs. This study focuses on strain engineering of homojunction TFETs, but it can be combined with heterojunction engineering and/or doping engineering to further boost the performances.


%



%
%

\ifCLASSOPTIONcaptionsoff
  \newpage
\fi

\end{document}